\documentclass[twocolumn,showpacs,preprintnumbers,amsmath,amssymb,prl,longbibliography,superscriptaddress]{revtex4-1}
\usepackage{bbm}
\usepackage{mathrsfs}
\usepackage{graphicx}
\usepackage{dcolumn}
\usepackage{bm}
\usepackage{amsmath}
\usepackage{amsfonts}
\usepackage{color}
\usepackage[colorlinks=true,citecolor=cyan,anchorcolor=blue]{hyperref}
\usepackage{floatrow}

\begin{document}

\title{Unoccupied Topological Surface State in MnBi$_2$Te$_4$}
\author{Yadong Jiang}
\affiliation{State Key Laboratory of Surface Physics and Department of Physics, Fudan University, Shanghai 200433, China}
\author{Zhaochen Liu}
\affiliation{State Key Laboratory of Surface Physics and Department of Physics, Fudan University, Shanghai 200433, China}
\author{Jing Wang}
\thanks{wjingphys@fudan.edu.cn}
\affiliation{State Key Laboratory of Surface Physics and Department of Physics, Fudan University, Shanghai 200433, China}
\affiliation{Institute for Nanoelectronic Devices and Quantum Computing, Fudan University, Shanghai 200433, China}
\affiliation{Zhangjiang Fudan International Innovation Center, Fudan University, Shanghai 201210, China}

\begin{abstract}
The unoccupied part of the band structure in the magnetic topological insulator MnBi$_2$Te$_4$ is studied by first-principles calculations. We find a second, unoccupied topological surface state with similar electronic structure to the celebrated occupied topological surface state. This state is energetically located approximate $1.6$~eV above the occupied Dirac surface state around $\Gamma$ point, which permit it to be directly observed by the two-photon angle-resolved photoemission spectroscopy. We propose a unified effective model for the occupied and unoccupied surface states. Due to the direct optical coupling between these two surface states, we further propose two optical effects to detect the unoccupied surface state. One is the polar Kerr effect in odd layer from nonvanishing ac Hall conductance $\sigma_{xy}(\omega)$, and the other is higher-order terahertz-sideband generation in even layer, where the non-vanishining Berry curvature of the unoccupied surface state is directly observed from the giant Faraday rotation of optical emission.
\end{abstract}

\date{\today}


\maketitle

\emph{Introduction.}
Topology has become one of the central topics in condensed matter physics~\cite{kane2005,kane2005b,bernevig2006c,koenig2007,hasan2010,qi2011,xiao2010,tokura2019,wang2017c}. A prime example is the topological insulators (TI) which is characterized by an insulating bulk and the hallmark Dirac surface states (SS) protected by the time-reversal symmetry~\cite{fu2007,xia2009,zhang2009,chen2009}. The linearly dispersing SS has a chiral spin texture with a nontrivial Berry phase, which is promising for spintronics applications~\cite{roushan2009,mellnik2014}. A number of interesting phenomena are associated with the symmetry-breaking of SS in TI, such as quantum anomalous Hall effect~\cite{qi2008,chang2013b,deng2020}, quantized Kerr rotation~\cite{maciejko2010,tse2010,wul2016,okada2016,dziom2017,mogi2021}, and image magnetic monopole~\cite{qi2009b} in magnetic TI, as well as Majorana fermion in the presence of superconductors~\cite{fu2008}.

Most of these phenomena are focused on the low energy electronic excitations. There are also quite a few work studying the excited electronic states and their dynamics in TI~\cite{hsieh2011,wangyh2012,niesner2012,sobota2013,patankar2015}. An interesting example is the discovery of unoccupied SS in Bi$_2$Se$_3$, which has similar electronic structure and physical origin to the occupied topological SS~\cite{niesner2012,sobota2013}. Recently an intrinsic magnetic TI MnBi$_2$Te$_4$ has been discovered~\cite{zhang2019,li2019,gong2019,otrokov2019,hao2019,lih2019,chen2019,swatek2020,yan2019}. The material consists of van der Waals coupled Te-Bi-Te-Mn-Te-Bi-Te septuple layers (SL) arranged along trigonal $z$ axis, so it can be viewed as layered TI Bi$_2$Te$_3$ with each of its Te-Bi-Te-Bi-Te quintuple layers intercalated by an additional Mn-Te bilayer. The resultant MnBi$_2$Te$_4$ remains a TI but now becomes intrinsically magnetic, where the hallmark first topological SS has been observed~\cite{gong2019,otrokov2019,hao2019,lih2019,chen2019,swatek2020}. The intimate relation between MnBi$_2$Te$_4$ and Bi$_2$Te$_3$ family motivates us to study the topological properties of higher-excited electronic states in MnBi$_2$Te$_4$.

In this paper, we investigate unoccupied part of the band structure in MnBi$_2$Te$_4$ by first-principles calculations. We reveal a second, unoccupied topological SS which has similar physical origin but is energetically located 1.6~eV above the well-known occupied SS. Moreover, we find that the direct optical coupling between the occupied and unoccupied SS permits it to be directly observed by the two-photon photoemission (2PPE) and higher-order terahertz-sideband generation (HSG).

\emph{Band structure and parity analysis.}
MnBi$_2$Te$_4$ has a rhombohedral crystal structure with space group $D_{3d}^5$ (No.~166). The magnetism originates from the Mn$^{2+}$ ions in the crystal. Below a N\'eel temperature of $T_N=25$~K, the system develops $A$-type antiferromagnetic (AFM) order with an out-of-plane easy axis, which is ferromagnetic (FM) within each SL but AFM between adjacent SL along $z$ axis~\cite{zhang2019,li2019,gong2019,otrokov2019}. The existence of inversion symmetry $\mathcal{P}$, with the Mn site as the inversion center, enables us to construct eigenstates with definite parity.

\begin{figure}[b]
\begin{center}
\includegraphics[width=3.4in, clip=true]{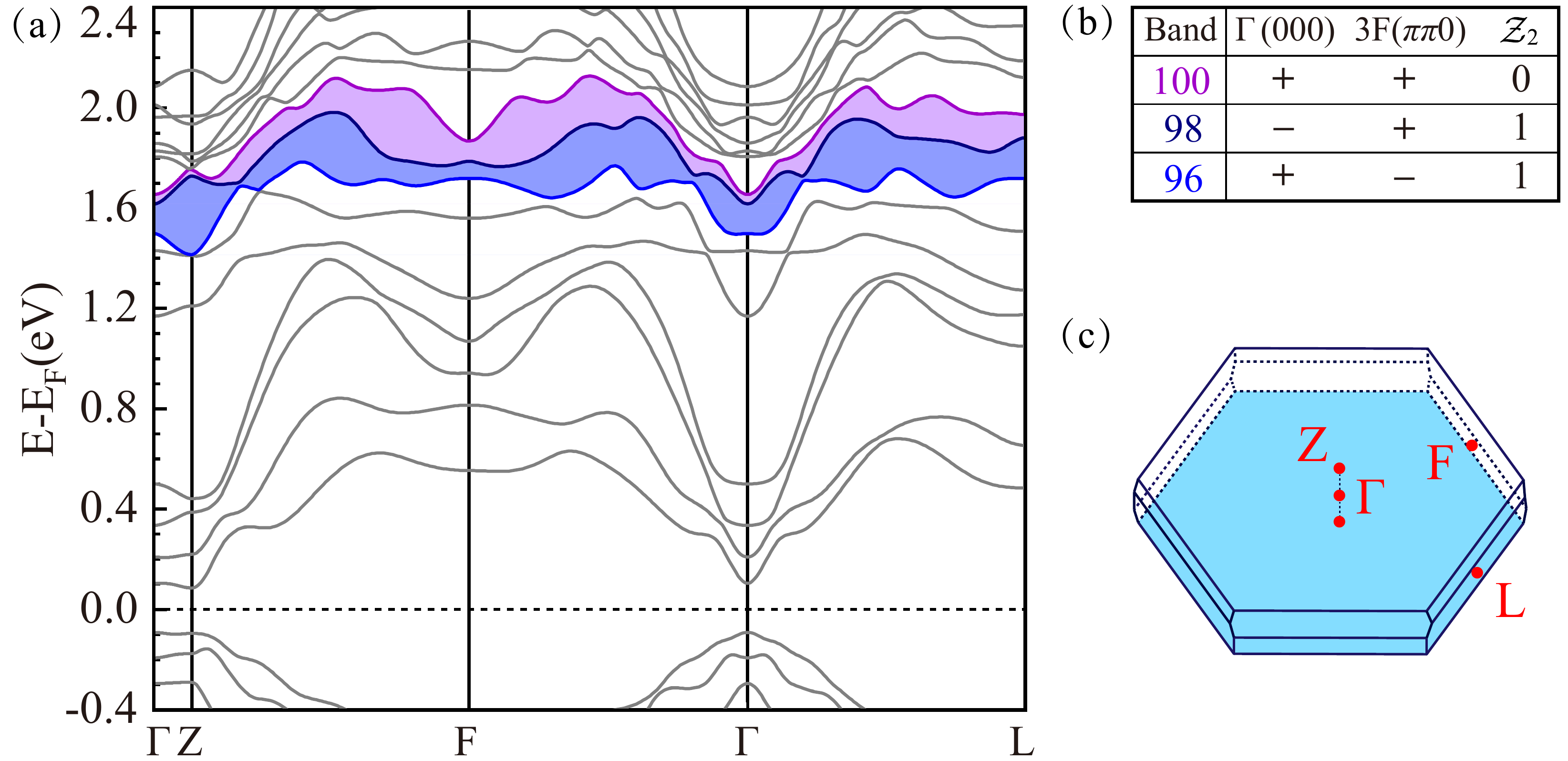}
\end{center}
\caption{(a) Electronic structure of AFM-$z$ ground state in MnBi$_2$Te$_4$. The topologically non-trivial unoccupied bands are colored. All energy bands have twofold degeneracy due to $\mathcal{S}$ and $\mathcal{P}$ symmetry. (b) The parity product of the unoccupied bands at the TRIM with $\bar{\mathbf{G}}\cdot\tau_{1/2}=n\pi$. (c) Brillouin zone. The four inequivalent TRIM are $\Gamma(000)$, $L(\pi00)$, $F(\pi\pi0)$, and $Z(\pi\pi\pi)$.}
\label{fig1}
\end{figure}

\begin{figure*}
\begin{center}
\includegraphics[width=\textwidth]{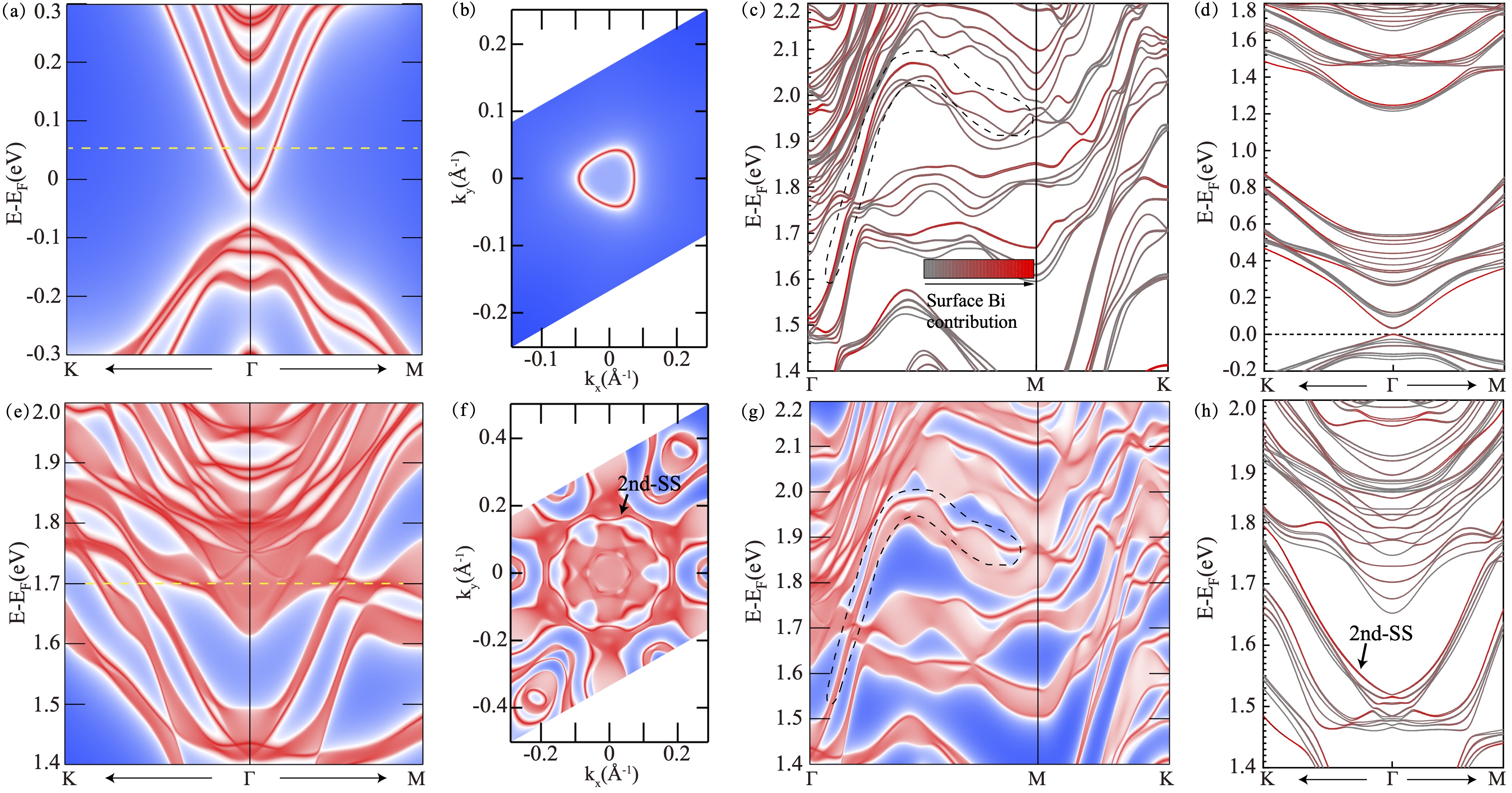}
\end{center}
\caption{Band structure calculations of SS in MnBi$_2$Te$_4$. (a), (e) \& (g) Energy and momentum dependence of the LDOS for AFM-$z$ state on the (111) surface. (b) \& (f) The constant energy contour for first and second SS at energy level $0.05$~eV and $1.7$~eV, respectively. (c) \& (d) Slab band structure calculation of 8 SL MnBi$_2$Te$_4$. The top of valence band is set as Fermi energy. The color of bands represents the degree of surface localization. The SS in (c) shows a one-to-one correspondence with that in (g) as highlighted in the dashed box. (h) is a magnified view of (d) with energy between $1.4$-$2.0$~eV.}
\label{fig2}
\end{figure*}

First-principles calculations are employed to investigate the electronic structure of MnBi$_2$Te$_4$, where the detailed methods can be found in the Supplemental Material~\cite{supple}. The calculations were based on the experimentally determined crystal structure~\cite{otrokov2019}. The AFM-$z$ ground state breaks the time-reversal symmetry $\Theta$, however, a combined symmetry $\mathcal{S}=\Theta\tau_{1/2}$ is preserved, where $\tau_{1/2}$ is the half translation operator connecting adjacent spin-up and -down Mn atomic layers. Here the operator $\mathcal{S}$ is antiunitary with $\mathcal{S}^2=-e^{-i\textbf{k}\cdot\tau_{1/2}}$, and $\mathcal{S}^2=-1$ on the Brillouin zone (BZ) plane $\textbf{k}\cdot\tau_{1/2}=0$. Therefore, similar to $\Theta$ in time-reversal invariant TI, $\mathcal{S}$ could also lead to a $\mathcal{Z}_2$ classification~\cite{mong2010}, where the topological invariant is only well defined on the BZ plane with $\mathbf{k}\cdot\tau_{1/2}=0$. Since $\mathcal{P}$ is preserved, the $\mathcal{Z}_2$ invariant is simply the parity of the wave functions of all occupied bands at time-reversal-invariant momenta (TRIM) in the BZ proposed by Fu and Kane~\cite{fu2007a}. Here we only need to consider four TRIM ($\Gamma$ and three $F$) with $\bar{\mathbf{G}}\cdot\tau_{1/2}=n\pi$.

The previous studies have revealed the nontrivial $\mathcal{Z}_2$ invariant for all occupied bands~\cite{zhang2019,li2019}, which signifies the 1st topological SS shown in Fig.~\ref{fig2}(a) and~\ref{fig2}(b). Here we focus on the topologically non-trivial unoccupied state above the Fermi energy. As shown in Fig.~\ref{fig1}(a), the different parity product between $\Gamma$ and $F$ leads to $\mathcal{Z}_2=1$, if we put the \emph{artificial} Fermi energy at the blue region (namely between No.~96th and 98th bands) or at the purple region (namely between No.~98th and 100th bands). Such nontrivial $\mathcal{Z}_2$ suggests the existence of unoccupied topological bands. The band inversion for the unoccupied topological states (No.~96th and 100th bands) happens between $p^{\pm}_{x,y}$ orbitals of Bi~\cite{supple}, which is slightly different from the occupied topological states where the band inversion is between $p^{+}_{z}$ orbital of Bi and $p^{-}_{z}$ orbital of Te. Therefore, the second unoccupied topologically non-trivial SS is expected to appear between 96th and 100th band, as shown by shade in Fig.~\ref{fig1}(a). Moreover, we find that both of the band structure and $\mathcal{Z}_2$ invariant are insensitive to the lattice constant ($\pm0.5\%$ strain) as well as the choice of different van der Waals interaction functional~\cite{supple}.

\emph{2nd SS.}
The existence of topological SS is the hallmark of TI. To explore the 1st and 2nd SS, we now turn to the local density of states (LDOS) and slab calculations. It is worth mentioning that the TI state in AFM MnBi$_2$Te$_4$ protected by $\mathcal{S}$ is topological in a weaker sense than the strong TI protected by $\Theta$, which manifests in that the existence of gapless SS depends on the surface plane. The 1st SS on the (111) surface is clearly shown in Fig.~\ref{fig2}(a) and~\ref{fig2}(b), which is gapped and accompanied by a triangular Fermi surface, for $\mathcal{S}$ is broken. While the 2nd SS on (111) surface is highlighted by arrow in Fig.~\ref{fig2}(e) and~\ref{fig2}(f). The 2nd SS is buried among the bulk state along $\Gamma$-$Z$ due to projection to surface $\bar{\Gamma}$ point in Fig.~\ref{fig2}(e). To better resolve the 2nd SS from the complex electronic structure, we further display the slab calculation of 8 SL in Fig.~\ref{fig2}(c) for comparison. One can identify an unambiguous one-to-one correspondence of the 2nd SS across the entire BZ between Fig.~\ref{fig2}(c) and Fig.~\ref{fig2}(g). This state is energetically located approximate $1.6\sim1.9$~eV above the occupied 1st topological SS around $\Gamma$ point, though the energy position shifts a little bit in Fig.~\ref{fig2}(c) and Fig.~\ref{fig2}(g). As expected, the 2nd SS is also magnetically gapped at $\Gamma$ from the magnified view in Fig.~\ref{fig2}(h). The magnetic gap of 2nd SS is smaller than that of the 1st SS, because the exchange field is from Mn$^{2+}$ $d$ orbitals which lie far below the Fermi level. Moreover, just like the 1st SS, the 2nd SS exists only in the presence of crystal spin-orbit coupling, as evidenced by calculations in Fig.~\ref{fig2} and Supplemental Material~\cite{supple}. This suggests that both SSs share the same physical origin, as they both arise due to band inversion of bulk states in the presence of strong spin-orbit coupling.

\emph{Effective model.}
We study the surface band structure near $\Gamma$ using $\mathbf{k}\cdot\mathbf{p}$ theory. To lowest order in $\mathbf{k}$, the $2\times2$ effective Hamiltonian reads $H_0=v(k_x\sigma_y-k_y\sigma_x)+m\sigma_z$, which describes an isotropic gapped 2D Dirac fermion. The Fermi surface of $H_0$ is a circle at any Fermi energy. Thus the anisotropic Fermi surface for 1st and 2nd SS in Fig.~\ref{fig2} can only be explained by higher order terms in the $\mathbf{k}\cdot\mathbf{p}$ Hamiltonian $\mathcal{H}(\mathbf{k})$ that breaks the emerging $U(1)$ rotational symmetry of $H_0$. From the constraint of crystal symmetry, $C_3$ around the trigonal $z$ axis transforms the momentum and spin as $C_3: k_\pm\rightarrow e^{\pm i2\pi/3}k_{\pm}$, $\sigma_\pm\rightarrow e^{\pm i2\pi/3}\sigma_\pm$, $\sigma_z\rightarrow\sigma_z$, where $k_\pm=k_x\pm ik_y$, $\sigma_\pm=\sigma_x\pm i\sigma_y$ and $k_x$ is in $\Gamma K$ direction. We then find that $\mathcal{H}(\mathbf{k})$ takes the following form up to third order in $\mathbf{k}$,
\begin{equation}\label{model}
\mathcal{H}_{i}(\mathbf{k})=E_0^{i}(\mathbf{k})+v^i_k(k_y\sigma_x-k_x\sigma_y)+\frac{\lambda_i}{2}(k_+^3+k_-^3)\sigma_z+\Delta_i\sigma_z,
\end{equation}
where $i=1,2$ denote the 1st and 2nd SS, respectively, $E_0^i(\mathbf{k})=\epsilon^i_0+k^2/2m^*_i$ generates particle-hole asymmetry, the Dirac velocity $v^i_k=v_i(1+\alpha_ik^2)$ has a second-order correction, $\lambda_i$ is the threefold warping term similar to Bi$_2$Te$_3$~\cite{fu2009}. $\Delta_i$ is the exchange field along $z$-axis on (111) surface introduced by magnetic ordering, which is odd under any mirror symmetry in two dimensions. The surface band dispersion of $\mathcal{H}_i(\mathbf{k})$ is
\begin{equation}\label{dispersion}
E_\pm^i(\mathbf{k})=E_0^i(\mathbf{k})\pm\sqrt{v_i^2k^2+(m_i+\lambda_i(k_x^3-3k_xk_y^2))^2}.
\end{equation}
Here $E_\pm$ denotes the upper and lower band. The shape of constant energy contour is energy-dependent and always forms a closed loop around $\Gamma$. Although the Hamiltonian $\mathcal{H}$ is threefold invariant, the band structure in Eq.~(\ref{dispersion}) is approximately sixfold symmetric when the energy is far away from the Dirac point, which is consistent with Fig.~\ref{fig2}(f). In the following we propose several optical experiments to reveal the properties of 2nd SS.

\emph{2PPE measurement.} To experimentally identify the 2nd unoccupied SS, a direct way is to employ angle-resolved 2PPE spectroscopy. Distinct from conventional one-photon photoemission for occupied states measurement, 2PPE could access unoccupied states~\cite{niesner2012,sobota2013}. In 2PPE process, a photon first excites an electron from below Fermi energy to an unoccupied intermediate state, and a second photon, further excites the electron above the vaccum.

\begin{figure}[t]
\begin{center} 
\includegraphics[width=3.4in,clip=true]{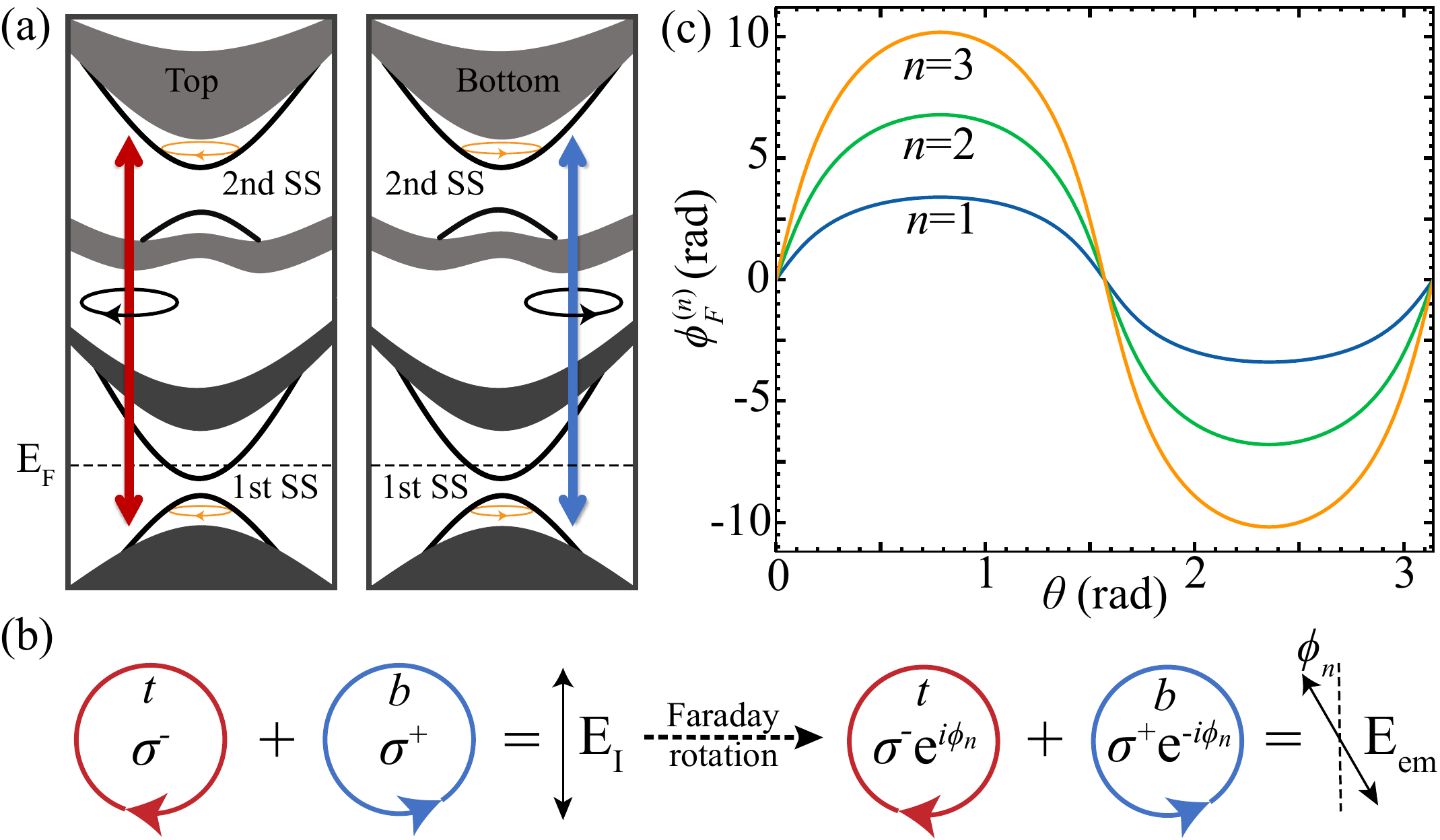}    
\end{center}
\caption{(a) Schematics of HSG between the lower band of 1st SS and the upper band of 2nd SS on $t$ and $b$ surfaces, where the optical selection rules are opposite. (b) The linear polarized light from two opposite circular polarization $|\sigma_-\rangle+|\sigma_+\rangle$ will excite optical transitions in two layers, resulting in the coherent superposition of electron-hole pair state $|t\rangle+|b\rangle$. The cyclic evolutions under THz field accumulate opposite Berry phase $e^{i\phi_n}|t\rangle+e^{-i\phi_n}|b\rangle$, which further leads to Faraday rotation of emission $e^{i\phi_n}|\sigma_-\rangle+e^{-i\phi_n}|\sigma_+\rangle$. (c) The Faraday rotation angle $\phi^{(n)}_F$ as function of the polarization angle $\theta$ of elliptical driving laser at $n=1,2,3$.}
\label{fig3}
\end{figure}

\emph{HSG in even SL.} 
In even SL MnBi$_2$Te$_4$, $\mathcal{P}\Theta$ is conserved due to full compensated magnetic layers. Therefore, the ac Hall conductance $\sigma_{xy}(\omega)$ vanishes, which further leads to vanishing linear optical Kerr or Faraday effect. In the limit of $\omega\rightarrow0$, $\sigma_{xy}(0)=0$ is one transport signature for axion insulator~\cite{wang2014a,wang2015b,mogi2017,xiao2018,liu2020,zhang2019}, where the interesting topological magnetoelectric effect associated with 1st SS is proposed~\cite{qi2008,wang2015b,nomura2011,yu2019,liuzc2020a,liuzc2020b}. Here we employ the extreme nonlinear optical phenomena namely HSG to detect the nontrivial Berry curvature of 2nd SS around $\Gamma$ point. HSG could have interesting effects due to nontrivial vacuum states of materials~\cite{liu2007,zaks2012,yang2013,yang2014,banks2017,langer2018}. Previous studies of the multi-valley system such as monolayer MoS$_2$ have shown that the finite valley Berry curvature is revealed through HSG~\cite{yang2013,yang2014,banks2017}. In MoS$_2$, two degenerate valleys related by $\Theta$ have opposite Berry curvature and optical selection rule~\cite{xiao2012}. The quantum trajectories of optically excited electron-hole pairs in two valleys driven under intense THz field accumulate opposite Berry phase, and the interference of optical transitions between two valleys leads to Faraday rotation of emission. 

Here we show that the optical transitions of the \emph{gapped} 1st SS to 2nd SS on the top and bottom surfaces in even SL is the same as that of two valleys in MoS$_2$. The effective Hamiltonian for even SL is $H_i=E_0^i+v_i(k_y\sigma_x-k_x\sigma_y)\tau_z+\Delta_i\sigma_z\tau_z$, with the basis of $|t\uparrow\rangle$, $|t\downarrow\rangle$, $|b\uparrow\rangle$, and $|b\downarrow\rangle$, where $t$ and $b$ denote the top and bottom surfaces and $\uparrow$ and $\downarrow$ represent spin up and down states, respectively. $\sigma_{x,y,z}$ and $\tau_z$ are Pauli matrices for spin and layers. $t$ and $b$ layers are decoupled. The Bloch state is denoted as $\psi_i^\pm(\mathbf{k})=e^{i\mathbf{k}\cdot\mathbf{r}}|i^\pm,\mathbf{k}\rangle$, where $i=1,2$ refer the 1st and 2nd SS, respectively, and $\pm$ denote upper and lower band, respectively. The optical transition from the lower band of 1st SS to upper band of 2nd SS is in the visible range as shown in Fig.~\ref{fig3}(a). The inter-band dipole moment is $\mathbf{d}_{t/b}(\mathbf{k})=-ie\left\langle\psi_2^+(\mathbf{k})\right|\nabla_{\mathbf{k}}\left|\psi^-_1(\mathbf{k})\right\rangle_{t/b}\approx d_{12}(\mathbf{e}_x\mp i\mathbf{e}_y)$, \emph{i.e.}, the SS on $t$ and $b$ can be optically pumped by opposite circular polarized light. The optical selection rules remain the same even with finite hybridization between these two surfaces~\cite{supple}. Also the Berry curvature for 2nd SS on $t$ and $b$ are opposite. The surface index in even SL magnetic TI is in exact analogy to valley index in MoS$_2$.

After the weak optical laser pumping $\mathbf{E}_{I}e^{-i\Omega t}$, the excited pairs of electron-hole reside at SS $|2^+,\mathbf{k}\rangle$ and $|1^-,\mathbf{k}\rangle$ states, repsectively. The Berry phase is accumulated by varying the parameter $\mathbf{k}$ in a closed path. Now in the presence of a strong THz driving field $\mathbf{F}(t)$, the minimal coupling leads to the time dependent Hamiltonian $H(\tilde{\mathbf{k}}(t))$ via $\tilde{\mathbf{k}}(t)\rightarrow \mathbf{k}+e\mathbf{A}(t)$, with $\mathbf{F}=-\partial\mathbf{A}/\partial t$. The instantaneous eigenstate is $H(\mathbf{\tilde{k}}(t))|\mu,\mathbf{\tilde{k}}(t)\rangle=E_{\mathbf{\tilde{k}}(t)}^\mu|\mu,\mathbf{\tilde{k}}(t)\rangle$, where $\mu=2^+,1^-$ is band index. Here the lifetime of a nonequilibrium population of 2nd SS is unknown and is beyond the scope of the current study. Since the 2nd SS and 1st SS are similar, and if we take the experiment value of the long-lived 1st SS in Bi$_2$Se$_3$ persisting $>10$~ps~\cite{sobota2012} for 2nd SS in MnBi$_2$Te$_4$, then the electron-hole pairs accelerated by the THz field could complete several cyclic evolutions in $\mathbf{k}$ space before they scatters into bulk states. Thus we assume the states evolve in $\mathbf{k}$-space would follow $\tilde{\mathbf{k}}(t)$ adiabatically. The linear optical response now is $\mathbf{P}(t)=i\int^{t}_{-\infty}dt'\int d\mathbf{k} \mathbf{d}^*_{\mathbf{\tilde{k}}(t)} \mathbf{d}_{\mathbf{\tilde{k}}(t')}\cdot \mathbf{E}_{I} e^{-i\int^{t}_{t'}\delta E_{\mathbf{\tilde{k}}(\tau)}d\tau+i\int^{t}_{t'}\mathcal{A}_{\mathbf{\tilde{k}}(\tau)}\cdot d\mathbf{\tilde{k}}(\tau)-i\Omega t'}$, where $\mathcal{A}_{\mathbf{\tilde{k}}}=\mathcal{A}^+_{\mathbf{\tilde{k}}}-\mathcal{A}^-_{\mathbf{\tilde{k}}}$ is the combined Berry connection between electron-hole pairs, $\mathcal{A}^{\mu}_{\tilde{\mathbf{k}}}=i\langle\mu,\tilde{\mathbf{k}}|\nabla_{\mathbf{k}}|\mu,\tilde{\mathbf{k}}\rangle$ is the Abelian Berry connection, $\mathbf{d}_{\mathbf{\tilde{k}}}=-ie\langle\psi_2^+(\tilde{\mathbf{k}})|\nabla_{\mathbf{k}}|\psi^-_1(\tilde{\mathbf{k}})\rangle$ is the instantaneous dipole moment, and $\delta E_{\mathbf{\tilde{k}}}=E^{2+}_{\mathbf{\tilde{k}}}-E^{1-}_{\mathbf{\tilde{k}}}$ is energy of electron-hole pair~\cite{yang2013}. Then for an elliptically polarized THz field $\mathbf{F}(t)=F(\cos\theta\cos(\omega t),\sin\theta\sin(\omega t),0)$, the optical response at $t_n=nT=2n\pi/\omega$ contains explicitly the Berry phase $\phi_F^{(n)}=\int^{t_n}_0\mathcal{A}_{\tilde{\mathbf{k}}(\tau)}\cdot d\tilde{\mathbf{k}}(\tau)$, where  the elliptical path $\tilde{\mathbf{k}}(t)=(k_x-k_0\cos\theta\sin(\omega t),k_y+k_0\sin\theta\cos(\omega t),k_z)$, $k_0=eF/\omega$.

Now if the linearly polarized laser $\mathbf{E}_I=E\delta(t)\mathbf{e}_{\parallel}$ is applied in the $x$-$y$ plane, then the dipole moment tensor $(\mathbf{d}_{\mathbf{k}}^*\mathbf{d}_{\mathbf{k}})^{t/b}=|d_{12}|^2(\mathbf{e}_x\mathbf{e}_x+\mathbf{e}_y\mathbf{e}_y\mp i(\mathbf{e}_{x}\mathbf{e}_{y}-\mathbf{e}_{y}\mathbf{e}_{x}))$, which leads to opposite Berry phases on $t$ and $b$ as illustrated in Fig.~\ref{fig3}(b). Thus the Faraday rotation of emission is $\phi^{(n)}_F(\theta)=n\int\mathcal{B}d\mathbf{k}^2$, where $\mathcal{B}=\sum_{i=1}^2v_i^2\Delta_i/(\Delta_i^2+v_i^2\mathbf{k}^2)^{3/2}$ is the Berry curvature. For an estimation, taking $v_1\approx v_2\approx0.25$~eV$\cdot$nm, $\Delta_1\approx0.03$~eV, $\Delta_2\approx0.01$~eV, $\omega=4$~meV, $F=8$~kV$\cdot$cm$^{-1}$, and $k_0=0.2$~nm$^{-1}$. Then the estimated Faraday rotation angle is $\phi_{1}\approx3.5$~rad as in shown Fig.~\ref{fig3}(c), which is surprisingly big due to giant Berry curvature from small Dirac gap (two orders of magnitude larger than that in MoS$_2$~\cite{yang2013}).

The Faraday rotation from the Berry phase is robust as protected by $\mathcal{P}\Theta$. Moreover, without the elliptical THz field, the Faraday rotation in this material has to vanish due to $\mathcal{P}\Theta$. The resonant optical transitions from $\mathbf{E}_I$ will inevitably involve other trivial bulk states, however, without nontrivial Berry curvature in topological bands, these optical processes will not contribute to Faraday rotation. All of these provide a sharp experimental signature for 2nd SS. The Berry phase will only be slightly changed by taking into account the trigonal warping term. The Faraday rotation is also robust against the decoherence of electron-hole pairs due to scattering. However, to observe this effect, the layer coherence time needs to be longer than the period of THz field. The SS on the two layers are physically decoupled, thus the layer coherence timescale is expected to be longer than that of electron-hole recombination.

\emph{Kerr effect in odd SL.} In odd SL MnBi$_2$Te$_4$, $\mathcal{P}\Theta$ is broken due to an uncompensated magnetic layer. Therefore, the ac Hall conductance $\sigma_{xy}(\omega)$ is nonvanishing. In the limit of $\omega\rightarrow0$, $\sigma_{xy}(0)=\pm e^2/h$ is the quantized anomalous Hall effect in transport~\cite{deng2020}. Now the finite frequency $\sigma_{xy}(\omega)$ leads to the polar Kerr effect. For a thin film that is much thinner than the optical wavelength, the Kerr rotation angle $\theta_K=(8\pi/c(n^2-1))\text{Re}(\sigma_{xy})$~\cite{supple}, where $c$ is velocity of light, $n$ is the refractive index of substrate. The unoccupied topological band associated with 2nd SS is a higher energy replica of the band inversion that takes place near the Fermi energy. Transitions resonant with spin-orbit avoided band crossings are shown to make a large contribution to $\text{Re}(\sigma_{xy})$~\cite{fang2005}. Therefore, similar to the resonant Kerr effect in magnetically doped TI~\cite{patankar2015}, a resonant Kerr angle $\theta_K$ should occur at $\omega\sim1.6$~eV in odd SL~\cite{supple}.
 
\emph{Conclusion.}
In summary, we present a study of the unoccupied part of the band structure in MnBi$_2$Te$_4$. A 2nd unoccupied topological SS is predicted, which can be directly observed by 2PPE spectroscopy. Several optical experiments are further proposed to reveal the 2nd SS. Interestingly, due to direct optical coupling between these two SS, a giant Faraday rotation induced by the Berry curvature in 2nd SS under strong THz fields can be observed in even SL through HSG. This proposal also applies to magnetic TI heterostructure~\cite{mogi2017,xiao2018}. We hope the theoretical work here can motivate the study of higher-excited states in vast topological materials.

\begin{acknowledgments}
This work is supported by the National Key Research Program of China under Grant No.~2019YFA0308404, the Natural Science Foundation of China through Grant No.~11774065, Shanghai Municipal Science and Technology Major Project under Grant No.~2019SHZDZX01, Science and Technology Commission of Shanghai Municipality under Grant No.~20JC1415900, and the Natural Science Foundation of Shanghai under Grant No.~19ZR1471400.
\end{acknowledgments}


\begin{thebibliography}{62}%
\makeatletter
\providecommand \@ifxundefined [1]{%
 \@ifx{#1\undefined}
}%
\providecommand \@ifnum [1]{%
 \ifnum #1\expandafter \@firstoftwo
 \else \expandafter \@secondoftwo
 \fi
}%
\providecommand \@ifx [1]{%
 \ifx #1\expandafter \@firstoftwo
 \else \expandafter \@secondoftwo
 \fi
}%
\providecommand \natexlab [1]{#1}%
\providecommand \enquote  [1]{``#1''}%
\providecommand \bibnamefont  [1]{#1}%
\providecommand \bibfnamefont [1]{#1}%
\providecommand \citenamefont [1]{#1}%
\providecommand \href@noop [0]{\@secondoftwo}%
\providecommand \href [0]{\begingroup \@sanitize@url \@href}%
\providecommand \@href[1]{\@@startlink{#1}\@@href}%
\providecommand \@@href[1]{\endgroup#1\@@endlink}%
\providecommand \@sanitize@url [0]{\catcode `\\12\catcode `\$12\catcode
  `\&12\catcode `\#12\catcode `\^12\catcode `\_12\catcode `\%12\relax}%
\providecommand \@@startlink[1]{}%
\providecommand \@@endlink[0]{}%
\providecommand \url  [0]{\begingroup\@sanitize@url \@url }%
\providecommand \@url [1]{\endgroup\@href {#1}{\urlprefix }}%
\providecommand \urlprefix  [0]{URL }%
\providecommand \Eprint [0]{\href }%
\providecommand \doibase [0]{http://dx.doi.org/}%
\providecommand \selectlanguage [0]{\@gobble}%
\providecommand \bibinfo  [0]{\@secondoftwo}%
\providecommand \bibfield  [0]{\@secondoftwo}%
\providecommand \translation [1]{[#1]}%
\providecommand \BibitemOpen [0]{}%
\providecommand \bibitemStop [0]{}%
\providecommand \bibitemNoStop [0]{.\EOS\space}%
\providecommand \EOS [0]{\spacefactor3000\relax}%
\providecommand \BibitemShut  [1]{\csname bibitem#1\endcsname}%
\let\auto@bib@innerbib\@empty
\bibitem [{\citenamefont {Kane}\ and\ \citenamefont
  {Mele}(2005{\natexlab{a}})}]{kane2005}%
  \BibitemOpen
  \bibfield  {author} {\bibinfo {author} {\bibfnamefont {C.~L.}\ \bibnamefont
  {Kane}}\ and\ \bibinfo {author} {\bibfnamefont {E.~J.}\ \bibnamefont
  {Mele}},\ }\bibfield  {title} {\enquote {\bibinfo {title} {${Z}_{2}$
  topological order and the quantum spin hall effect},}\ }\href {\doibase
  10.1103/PhysRevLett.95.146802} {\bibfield  {journal} {\bibinfo  {journal}
  {Phys. Rev. Lett.}\ }\textbf {\bibinfo {volume} {95}},\ \bibinfo {pages}
  {146802} (\bibinfo {year} {2005}{\natexlab{a}})}\BibitemShut {NoStop}%
\bibitem [{\citenamefont {Kane}\ and\ \citenamefont
  {Mele}(2005{\natexlab{b}})}]{kane2005b}%
  \BibitemOpen
  \bibfield  {author} {\bibinfo {author} {\bibfnamefont {C.~L.}\ \bibnamefont
  {Kane}}\ and\ \bibinfo {author} {\bibfnamefont {E.~J.}\ \bibnamefont
  {Mele}},\ }\bibfield  {title} {\enquote {\bibinfo {title} {{Quantum Spin Hall
  Effect in Graphene}},}\ }\href {\doibase 10.1103/PhysRevLett.95.226801}
  {\bibfield  {journal} {\bibinfo  {journal} {Phys. Rev. Lett.}\ }\textbf
  {\bibinfo {volume} {95}},\ \bibinfo {pages} {226801} (\bibinfo {year}
  {2005}{\natexlab{b}})}\BibitemShut {NoStop}%
\bibitem [{\citenamefont {Bernevig}\ \emph {et~al.}(2006)\citenamefont
  {Bernevig}, \citenamefont {Hughes},\ and\ \citenamefont
  {Zhang}}]{bernevig2006c}%
  \BibitemOpen
  \bibfield  {author} {\bibinfo {author} {\bibfnamefont {B.~Andrei}\
  \bibnamefont {Bernevig}}, \bibinfo {author} {\bibfnamefont {Taylor~L.}\
  \bibnamefont {Hughes}}, \ and\ \bibinfo {author} {\bibfnamefont {Shou-Cheng}\
  \bibnamefont {Zhang}},\ }\bibfield  {title} {\enquote {\bibinfo {title}
  {Quantum spin hall effect and topological phase transition in hgte quantum
  wells},}\ }\href {\doibase 10.1126/science.1133734} {\bibfield  {journal}
  {\bibinfo  {journal} {Science}\ }\textbf {\bibinfo {volume} {314}},\ \bibinfo
  {pages} {1757--1761} (\bibinfo {year} {2006})}\BibitemShut {NoStop}%
\bibitem [{\citenamefont {K\"onig}\ \emph {et~al.}(2007)\citenamefont
  {K\"onig}, \citenamefont {Wiedmann}, \citenamefont {Br\"une}, \citenamefont
  {Roth}, \citenamefont {Buhmann}, \citenamefont {Molenkamp}, \citenamefont
  {Qi},\ and\ \citenamefont {Zhang}}]{koenig2007}%
  \BibitemOpen
  \bibfield  {author} {\bibinfo {author} {\bibfnamefont {Markus}\ \bibnamefont
  {K\"onig}}, \bibinfo {author} {\bibfnamefont {Steffen}\ \bibnamefont
  {Wiedmann}}, \bibinfo {author} {\bibfnamefont {Christoph}\ \bibnamefont
  {Br\"une}}, \bibinfo {author} {\bibfnamefont {Andreas}\ \bibnamefont {Roth}},
  \bibinfo {author} {\bibfnamefont {Hartmut}\ \bibnamefont {Buhmann}}, \bibinfo
  {author} {\bibfnamefont {Laurens}\ \bibnamefont {Molenkamp}}, \bibinfo
  {author} {\bibfnamefont {Xiao-Liang}\ \bibnamefont {Qi}}, \ and\ \bibinfo
  {author} {\bibfnamefont {Shou-Cheng}\ \bibnamefont {Zhang}},\ }\bibfield
  {title} {\enquote {\bibinfo {title} {{Quantum Spin Hall Insulator State in
  HgTe Quantum Wells}},}\ }\href {\doibase 10.1126/science.1148047} {\bibfield
  {journal} {\bibinfo  {journal} {Science}\ }\textbf {\bibinfo {volume}
  {318}},\ \bibinfo {pages} {766--770} (\bibinfo {year} {2007})}\BibitemShut
  {NoStop}%
\bibitem [{\citenamefont {Hasan}\ and\ \citenamefont {Kane}(2010)}]{hasan2010}%
  \BibitemOpen
  \bibfield  {author} {\bibinfo {author} {\bibfnamefont {M.~Z.}\ \bibnamefont
  {Hasan}}\ and\ \bibinfo {author} {\bibfnamefont {C.~L.}\ \bibnamefont
  {Kane}},\ }\bibfield  {title} {\enquote {\bibinfo {title}
  {\textit{Colloquium}: Topological insulators},}\ }\href {\doibase
  10.1103/RevModPhys.82.3045} {\bibfield  {journal} {\bibinfo  {journal} {Rev.
  Mod. Phys.}\ }\textbf {\bibinfo {volume} {82}},\ \bibinfo {pages}
  {3045--3067} (\bibinfo {year} {2010})}\BibitemShut {NoStop}%
\bibitem [{\citenamefont {Qi}\ and\ \citenamefont {Zhang}(2011)}]{qi2011}%
  \BibitemOpen
  \bibfield  {author} {\bibinfo {author} {\bibfnamefont {Xiao-Liang}\
  \bibnamefont {Qi}}\ and\ \bibinfo {author} {\bibfnamefont {Shou-Cheng}\
  \bibnamefont {Zhang}},\ }\bibfield  {title} {\enquote {\bibinfo {title}
  {Topological insulators and superconductors},}\ }\href {\doibase
  10.1103/RevModPhys.83.1057} {\bibfield  {journal} {\bibinfo  {journal} {Rev.
  Mod. Phys.}\ }\textbf {\bibinfo {volume} {83}},\ \bibinfo {pages}
  {1057--1110} (\bibinfo {year} {2011})}\BibitemShut {NoStop}%
\bibitem [{\citenamefont {Xiao}\ \emph {et~al.}(2010)\citenamefont {Xiao},
  \citenamefont {Chang},\ and\ \citenamefont {Niu}}]{xiao2010}%
  \BibitemOpen
  \bibfield  {author} {\bibinfo {author} {\bibfnamefont {Di}~\bibnamefont
  {Xiao}}, \bibinfo {author} {\bibfnamefont {Ming-Che}\ \bibnamefont {Chang}},
  \ and\ \bibinfo {author} {\bibfnamefont {Qian}\ \bibnamefont {Niu}},\
  }\bibfield  {title} {\enquote {\bibinfo {title} {Berry phase effects on
  electronic properties},}\ }\href {\doibase 10.1103/RevModPhys.82.1959}
  {\bibfield  {journal} {\bibinfo  {journal} {Rev. Mod. Phys.}\ }\textbf
  {\bibinfo {volume} {82}},\ \bibinfo {pages} {1959--2007} (\bibinfo {year}
  {2010})}\BibitemShut {NoStop}%
\bibitem [{\citenamefont {Tokura}\ \emph {et~al.}(2019)\citenamefont {Tokura},
  \citenamefont {Yasuda},\ and\ \citenamefont {Tsukazaki}}]{tokura2019}%
  \BibitemOpen
  \bibfield  {author} {\bibinfo {author} {\bibfnamefont {Yoshinori}\
  \bibnamefont {Tokura}}, \bibinfo {author} {\bibfnamefont {Kenji}\
  \bibnamefont {Yasuda}}, \ and\ \bibinfo {author} {\bibfnamefont {Atsushi}\
  \bibnamefont {Tsukazaki}},\ }\bibfield  {title} {\enquote {\bibinfo {title}
  {Magnetic topological insulators},}\ }\href {\doibase
  10.1038/s42254-018-0011-5} {\bibfield  {journal} {\bibinfo  {journal} {Nat.
  Rev. Phys.}\ }\textbf {\bibinfo {volume} {1}},\ \bibinfo {pages} {126--143}
  (\bibinfo {year} {2019})}\BibitemShut {NoStop}%
\bibitem [{\citenamefont {Wang}\ and\ \citenamefont {Zhang}(2017)}]{wang2017c}%
  \BibitemOpen
  \bibfield  {author} {\bibinfo {author} {\bibfnamefont {Jing}\ \bibnamefont
  {Wang}}\ and\ \bibinfo {author} {\bibfnamefont {Shou-Cheng}\ \bibnamefont
  {Zhang}},\ }\bibfield  {title} {\enquote {\bibinfo {title} {Topological
  states of condensed matter},}\ }\href {\doibase 10.1038/NMAT5012} {\bibfield
  {journal} {\bibinfo  {journal} {Nature Mat.}\ }\textbf {\bibinfo {volume}
  {16}},\ \bibinfo {pages} {1062--1067} (\bibinfo {year} {2017})}\BibitemShut
  {NoStop}%
\bibitem [{\citenamefont {Fu}\ \emph {et~al.}(2007)\citenamefont {Fu},
  \citenamefont {Kane},\ and\ \citenamefont {Mele}}]{fu2007}%
  \BibitemOpen
  \bibfield  {author} {\bibinfo {author} {\bibfnamefont {Liang}\ \bibnamefont
  {Fu}}, \bibinfo {author} {\bibfnamefont {C.~L.}\ \bibnamefont {Kane}}, \ and\
  \bibinfo {author} {\bibfnamefont {E.~J.}\ \bibnamefont {Mele}},\ }\bibfield
  {title} {\enquote {\bibinfo {title} {Topological insulators in three
  dimensions},}\ }\href {\doibase 10.1103/PhysRevLett.98.106803} {\bibfield
  {journal} {\bibinfo  {journal} {Phys. Rev. Lett.}\ }\textbf {\bibinfo
  {volume} {98}},\ \bibinfo {pages} {106803} (\bibinfo {year}
  {2007})}\BibitemShut {NoStop}%
\bibitem [{\citenamefont {Xia}\ \emph {et~al.}(2009)\citenamefont {Xia},
  \citenamefont {Qian}, \citenamefont {Hsieh}, \citenamefont {Wray},
  \citenamefont {Pal}, \citenamefont {Lin}, \citenamefont {Bansil},
  \citenamefont {Grauer}, \citenamefont {Hor}, \citenamefont {Cava},\ and\
  \citenamefont {Hasan}}]{xia2009}%
  \BibitemOpen
  \bibfield  {author} {\bibinfo {author} {\bibfnamefont {Y.}~\bibnamefont
  {Xia}}, \bibinfo {author} {\bibfnamefont {D.}~\bibnamefont {Qian}}, \bibinfo
  {author} {\bibfnamefont {D.}~\bibnamefont {Hsieh}}, \bibinfo {author}
  {\bibfnamefont {L.}~\bibnamefont {Wray}}, \bibinfo {author} {\bibfnamefont
  {A.}~\bibnamefont {Pal}}, \bibinfo {author} {\bibfnamefont {H.}~\bibnamefont
  {Lin}}, \bibinfo {author} {\bibfnamefont {A.}~\bibnamefont {Bansil}},
  \bibinfo {author} {\bibfnamefont {D.}~\bibnamefont {Grauer}}, \bibinfo
  {author} {\bibfnamefont {Y.~S.}\ \bibnamefont {Hor}}, \bibinfo {author}
  {\bibfnamefont {R.~J.}\ \bibnamefont {Cava}}, \ and\ \bibinfo {author}
  {\bibfnamefont {M.~Z.}\ \bibnamefont {Hasan}},\ }\bibfield  {title} {\enquote
  {\bibinfo {title} {{Observation of a large-gap topological-insulator class
  with a single Dirac cone on the surface}},}\ }\href {\doibase
  10.1038/nphys1274} {\bibfield  {journal} {\bibinfo  {journal} {Nature Phys.}\
  }\textbf {\bibinfo {volume} {5}},\ \bibinfo {pages} {398--402} (\bibinfo
  {year} {2009})}\BibitemShut {NoStop}%
\bibitem [{\citenamefont {Zhang}\ \emph {et~al.}(2009)\citenamefont {Zhang},
  \citenamefont {Liu}, \citenamefont {Qi}, \citenamefont {Dai}, \citenamefont
  {Fang},\ and\ \citenamefont {Zhang}}]{zhang2009}%
  \BibitemOpen
  \bibfield  {author} {\bibinfo {author} {\bibfnamefont {Haijun}\ \bibnamefont
  {Zhang}}, \bibinfo {author} {\bibfnamefont {Chao-Xing}\ \bibnamefont {Liu}},
  \bibinfo {author} {\bibfnamefont {Xiao-Liang}\ \bibnamefont {Qi}}, \bibinfo
  {author} {\bibfnamefont {Xi}~\bibnamefont {Dai}}, \bibinfo {author}
  {\bibfnamefont {Zhong}\ \bibnamefont {Fang}}, \ and\ \bibinfo {author}
  {\bibfnamefont {Shou-Cheng}\ \bibnamefont {Zhang}},\ }\bibfield  {title}
  {\enquote {\bibinfo {title} {{Topological insulators in
  {$\mathrm{Bi_2Se_3}$}, {$\mathrm{Bi_2Te_3}$} and {$\mathrm{Sb_2Te_3}$} with a
  single Dirac cone on the surface}},}\ }\href {\doibase 10.1038/nphys1270}
  {\bibfield  {journal} {\bibinfo  {journal} {Nature Phys.}\ }\textbf {\bibinfo
  {volume} {5}},\ \bibinfo {pages} {438} (\bibinfo {year} {2009})}\BibitemShut
  {NoStop}%
\bibitem [{\citenamefont {Chen}\ \emph {et~al.}(2009)\citenamefont {Chen},
  \citenamefont {Analytis}, \citenamefont {Chu}, \citenamefont {Liu},
  \citenamefont {Mo}, \citenamefont {Qi}, \citenamefont {Zhang}, \citenamefont
  {Lu}, \citenamefont {Dai}, \citenamefont {Fang}, \citenamefont {Zhang},
  \citenamefont {Fisher}, \citenamefont {Hussain},\ and\ \citenamefont
  {Shen}}]{chen2009}%
  \BibitemOpen
  \bibfield  {author} {\bibinfo {author} {\bibfnamefont {Y.~L.}\ \bibnamefont
  {Chen}}, \bibinfo {author} {\bibfnamefont {J.~G.}\ \bibnamefont {Analytis}},
  \bibinfo {author} {\bibfnamefont {J.-H.}\ \bibnamefont {Chu}}, \bibinfo
  {author} {\bibfnamefont {Z.~K.}\ \bibnamefont {Liu}}, \bibinfo {author}
  {\bibfnamefont {S.-K.}\ \bibnamefont {Mo}}, \bibinfo {author} {\bibfnamefont
  {X.~L.}\ \bibnamefont {Qi}}, \bibinfo {author} {\bibfnamefont {H.~J.}\
  \bibnamefont {Zhang}}, \bibinfo {author} {\bibfnamefont {D.~H.}\ \bibnamefont
  {Lu}}, \bibinfo {author} {\bibfnamefont {X.}~\bibnamefont {Dai}}, \bibinfo
  {author} {\bibfnamefont {Z.}~\bibnamefont {Fang}}, \bibinfo {author}
  {\bibfnamefont {S.~C.}\ \bibnamefont {Zhang}}, \bibinfo {author}
  {\bibfnamefont {I.~R.}\ \bibnamefont {Fisher}}, \bibinfo {author}
  {\bibfnamefont {Z.}~\bibnamefont {Hussain}}, \ and\ \bibinfo {author}
  {\bibfnamefont {Z.-X.}\ \bibnamefont {Shen}},\ }\bibfield  {title} {\enquote
  {\bibinfo {title} {Experimental realization of a three-dimensional
  topological insulator, bi2te3},}\ }\href {\doibase 10.1126/science.1173034}
  {\bibfield  {journal} {\bibinfo  {journal} {Science}\ }\textbf {\bibinfo
  {volume} {325}},\ \bibinfo {pages} {178--181} (\bibinfo {year}
  {2009})}\BibitemShut {NoStop}%
\bibitem [{\citenamefont {Roushan}\ \emph {et~al.}(2009)\citenamefont
  {Roushan}, \citenamefont {Seo}, \citenamefont {Parker}, \citenamefont {Hor},
  \citenamefont {Hsieh}, \citenamefont {Qian}, \citenamefont {Richardella},
  \citenamefont {Hasan}, \citenamefont {Cava},\ and\ \citenamefont
  {Yazdani}}]{roushan2009}%
  \BibitemOpen
  \bibfield  {author} {\bibinfo {author} {\bibfnamefont {Pedram}\ \bibnamefont
  {Roushan}}, \bibinfo {author} {\bibfnamefont {Jungpil}\ \bibnamefont {Seo}},
  \bibinfo {author} {\bibfnamefont {Colin~V.}\ \bibnamefont {Parker}}, \bibinfo
  {author} {\bibfnamefont {Y.~S.}\ \bibnamefont {Hor}}, \bibinfo {author}
  {\bibfnamefont {D.}~\bibnamefont {Hsieh}}, \bibinfo {author} {\bibfnamefont
  {Dong}\ \bibnamefont {Qian}}, \bibinfo {author} {\bibfnamefont {Anthony}\
  \bibnamefont {Richardella}}, \bibinfo {author} {\bibfnamefont {M.~Z.}\
  \bibnamefont {Hasan}}, \bibinfo {author} {\bibfnamefont {R.~J.}\ \bibnamefont
  {Cava}}, \ and\ \bibinfo {author} {\bibfnamefont {Ali}\ \bibnamefont
  {Yazdani}},\ }\bibfield  {title} {\enquote {\bibinfo {title} {{Topological
  surface states protected from backscattering by chiral spin texture}},}\
  }\href {\doibase 10.1038/nature08308} {\bibfield  {journal} {\bibinfo
  {journal} {Nature}\ }\textbf {\bibinfo {volume} {460}},\ \bibinfo {pages}
  {1106--1109} (\bibinfo {year} {2009})}\BibitemShut {NoStop}%
\bibitem [{\citenamefont {Mellnik}\ \emph {et~al.}(2014)\citenamefont
  {Mellnik}, \citenamefont {Lee}, \citenamefont {Richardella}, \citenamefont
  {Grab}, \citenamefont {Mintun}, \citenamefont {Fischer}, \citenamefont
  {Vaezi}, \citenamefont {Manchon}, \citenamefont {Kim}, \citenamefont
  {Samarth},\ and\ \citenamefont {Ralph}}]{mellnik2014}%
  \BibitemOpen
  \bibfield  {author} {\bibinfo {author} {\bibfnamefont {A.~R.}\ \bibnamefont
  {Mellnik}}, \bibinfo {author} {\bibfnamefont {J.~S.}\ \bibnamefont {Lee}},
  \bibinfo {author} {\bibfnamefont {A.}~\bibnamefont {Richardella}}, \bibinfo
  {author} {\bibfnamefont {J.~L.}\ \bibnamefont {Grab}}, \bibinfo {author}
  {\bibfnamefont {P.~J.}\ \bibnamefont {Mintun}}, \bibinfo {author}
  {\bibfnamefont {M.~H.}\ \bibnamefont {Fischer}}, \bibinfo {author}
  {\bibfnamefont {A.}~\bibnamefont {Vaezi}}, \bibinfo {author} {\bibfnamefont
  {A.}~\bibnamefont {Manchon}}, \bibinfo {author} {\bibfnamefont {E.-A.}\
  \bibnamefont {Kim}}, \bibinfo {author} {\bibfnamefont {N.}~\bibnamefont
  {Samarth}}, \ and\ \bibinfo {author} {\bibfnamefont {D.~C.}\ \bibnamefont
  {Ralph}},\ }\bibfield  {title} {\enquote {\bibinfo {title} {{Spin-transfer
  torque generated by a topological insulator}},}\ }\href {\doibase
  10.1038/nature13534} {\bibfield  {journal} {\bibinfo  {journal} {Nature}\
  }\textbf {\bibinfo {volume} {511}},\ \bibinfo {pages} {449--451} (\bibinfo
  {year} {2014})}\BibitemShut {NoStop}%
\bibitem [{\citenamefont {Qi}\ \emph {et~al.}(2008)\citenamefont {Qi},
  \citenamefont {Hughes},\ and\ \citenamefont {Zhang}}]{qi2008}%
  \BibitemOpen
  \bibfield  {author} {\bibinfo {author} {\bibfnamefont {Xiao-Liang}\
  \bibnamefont {Qi}}, \bibinfo {author} {\bibfnamefont {Taylor~L.}\
  \bibnamefont {Hughes}}, \ and\ \bibinfo {author} {\bibfnamefont {Shou-Cheng}\
  \bibnamefont {Zhang}},\ }\bibfield  {title} {\enquote {\bibinfo {title}
  {Topological field theory of time-reversal invariant insulators},}\ }\href
  {\doibase 10.1103/PhysRevB.78.195424} {\bibfield  {journal} {\bibinfo
  {journal} {Phys. Rev. B}\ }\textbf {\bibinfo {volume} {78}},\ \bibinfo
  {pages} {195424} (\bibinfo {year} {2008})}\BibitemShut {NoStop}%
\bibitem [{\citenamefont {Chang}\ \emph {et~al.}(2013)\citenamefont {Chang},
  \citenamefont {Zhang}, \citenamefont {Feng}, \citenamefont {Shen},
  \citenamefont {Zhang}, \citenamefont {Guo}, \citenamefont {Li}, \citenamefont
  {Ou}, \citenamefont {Wei}, \citenamefont {Wang}, \citenamefont {Ji},
  \citenamefont {Feng}, \citenamefont {Ji}, \citenamefont {Chen}, \citenamefont
  {Jia}, \citenamefont {Dai}, \citenamefont {Fang}, \citenamefont {Zhang},
  \citenamefont {He}, \citenamefont {Wang}, \citenamefont {Lu}, \citenamefont
  {Ma},\ and\ \citenamefont {Xue}}]{chang2013b}%
  \BibitemOpen
  \bibfield  {author} {\bibinfo {author} {\bibfnamefont {Cui-Zu}\ \bibnamefont
  {Chang}}, \bibinfo {author} {\bibfnamefont {Jinsong}\ \bibnamefont {Zhang}},
  \bibinfo {author} {\bibfnamefont {Xiao}\ \bibnamefont {Feng}}, \bibinfo
  {author} {\bibfnamefont {Jie}\ \bibnamefont {Shen}}, \bibinfo {author}
  {\bibfnamefont {Zuocheng}\ \bibnamefont {Zhang}}, \bibinfo {author}
  {\bibfnamefont {Minghua}\ \bibnamefont {Guo}}, \bibinfo {author}
  {\bibfnamefont {Kang}\ \bibnamefont {Li}}, \bibinfo {author} {\bibfnamefont
  {Yunbo}\ \bibnamefont {Ou}}, \bibinfo {author} {\bibfnamefont {Pang}\
  \bibnamefont {Wei}}, \bibinfo {author} {\bibfnamefont {Li-Li}\ \bibnamefont
  {Wang}}, \bibinfo {author} {\bibfnamefont {Zhong-Qing}\ \bibnamefont {Ji}},
  \bibinfo {author} {\bibfnamefont {Yang}\ \bibnamefont {Feng}}, \bibinfo
  {author} {\bibfnamefont {Shuaihua}\ \bibnamefont {Ji}}, \bibinfo {author}
  {\bibfnamefont {Xi}~\bibnamefont {Chen}}, \bibinfo {author} {\bibfnamefont
  {Jinfeng}\ \bibnamefont {Jia}}, \bibinfo {author} {\bibfnamefont
  {Xi}~\bibnamefont {Dai}}, \bibinfo {author} {\bibfnamefont {Zhong}\
  \bibnamefont {Fang}}, \bibinfo {author} {\bibfnamefont {Shou-Cheng}\
  \bibnamefont {Zhang}}, \bibinfo {author} {\bibfnamefont {Ke}~\bibnamefont
  {He}}, \bibinfo {author} {\bibfnamefont {Yayu}\ \bibnamefont {Wang}},
  \bibinfo {author} {\bibfnamefont {Li}~\bibnamefont {Lu}}, \bibinfo {author}
  {\bibfnamefont {Xu-Cun}\ \bibnamefont {Ma}}, \ and\ \bibinfo {author}
  {\bibfnamefont {Qi-Kun}\ \bibnamefont {Xue}},\ }\bibfield  {title} {\enquote
  {\bibinfo {title} {{Experimental Observation of the Quantum Anomalous Hall
  Effect in a Magnetic Topological Insulator}},}\ }\href {\doibase
  10.1126/science.1234414} {\bibfield  {journal} {\bibinfo  {journal}
  {Science}\ }\textbf {\bibinfo {volume} {340}},\ \bibinfo {pages} {167--170}
  (\bibinfo {year} {2013})}\BibitemShut {NoStop}%
\bibitem [{\citenamefont {Deng}\ \emph {et~al.}(2020)\citenamefont {Deng},
  \citenamefont {Yu}, \citenamefont {Shi}, \citenamefont {Guo}, \citenamefont
  {Xu}, \citenamefont {Wang}, \citenamefont {Chen},\ and\ \citenamefont
  {Zhang}}]{deng2020}%
  \BibitemOpen
  \bibfield  {author} {\bibinfo {author} {\bibfnamefont {Yujun}\ \bibnamefont
  {Deng}}, \bibinfo {author} {\bibfnamefont {Yijun}\ \bibnamefont {Yu}},
  \bibinfo {author} {\bibfnamefont {Meng~Zhu}\ \bibnamefont {Shi}}, \bibinfo
  {author} {\bibfnamefont {Zhongxun}\ \bibnamefont {Guo}}, \bibinfo {author}
  {\bibfnamefont {Zihan}\ \bibnamefont {Xu}}, \bibinfo {author} {\bibfnamefont
  {Jing}\ \bibnamefont {Wang}}, \bibinfo {author} {\bibfnamefont {Xian~Hui}\
  \bibnamefont {Chen}}, \ and\ \bibinfo {author} {\bibfnamefont {Yuanbo}\
  \bibnamefont {Zhang}},\ }\bibfield  {title} {\enquote {\bibinfo {title}
  {Quantum anomalous hall effect in intrinsic magnetic topological insulator
  mnbi2te4},}\ }\href {\doibase 10.1126/science.aax8156} {\bibfield  {journal}
  {\bibinfo  {journal} {Science}\ }\textbf {\bibinfo {volume} {367}},\ \bibinfo
  {pages} {895--900} (\bibinfo {year} {2020})}\BibitemShut {NoStop}%
\bibitem [{\citenamefont {Maciejko}\ \emph {et~al.}(2010)\citenamefont
  {Maciejko}, \citenamefont {Qi}, \citenamefont {Drew},\ and\ \citenamefont
  {Zhang}}]{maciejko2010}%
  \BibitemOpen
  \bibfield  {author} {\bibinfo {author} {\bibfnamefont {Joseph}\ \bibnamefont
  {Maciejko}}, \bibinfo {author} {\bibfnamefont {Xiao-Liang}\ \bibnamefont
  {Qi}}, \bibinfo {author} {\bibfnamefont {H.~Dennis}\ \bibnamefont {Drew}}, \
  and\ \bibinfo {author} {\bibfnamefont {Shou-Cheng}\ \bibnamefont {Zhang}},\
  }\bibfield  {title} {\enquote {\bibinfo {title} {Topological quantization in
  units of the fine structure constant},}\ }\href {\doibase
  10.1103/PhysRevLett.105.166803} {\bibfield  {journal} {\bibinfo  {journal}
  {Phys. Rev. Lett.}\ }\textbf {\bibinfo {volume} {105}},\ \bibinfo {pages}
  {166803} (\bibinfo {year} {2010})}\BibitemShut {NoStop}%
\bibitem [{\citenamefont {Tse}\ and\ \citenamefont
  {MacDonald}(2010)}]{tse2010}%
  \BibitemOpen
  \bibfield  {author} {\bibinfo {author} {\bibfnamefont {Wang-Kong}\
  \bibnamefont {Tse}}\ and\ \bibinfo {author} {\bibfnamefont {A.~H.}\
  \bibnamefont {MacDonald}},\ }\bibfield  {title} {\enquote {\bibinfo {title}
  {Giant magneto-optical kerr effect and universal faraday effect in thin-film
  topological insulators},}\ }\href {\doibase 10.1103/PhysRevLett.105.057401}
  {\bibfield  {journal} {\bibinfo  {journal} {Phys. Rev. Lett.}\ }\textbf
  {\bibinfo {volume} {105}},\ \bibinfo {pages} {057401} (\bibinfo {year}
  {2010})}\BibitemShut {NoStop}%
\bibitem [{\citenamefont {Wu}\ \emph {et~al.}(2016)\citenamefont {Wu},
  \citenamefont {Salehi}, \citenamefont {Koirala}, \citenamefont {Moon},
  \citenamefont {Oh},\ and\ \citenamefont {Armitage}}]{wul2016}%
  \BibitemOpen
  \bibfield  {author} {\bibinfo {author} {\bibfnamefont {Liang}\ \bibnamefont
  {Wu}}, \bibinfo {author} {\bibfnamefont {M.}~\bibnamefont {Salehi}}, \bibinfo
  {author} {\bibfnamefont {N.}~\bibnamefont {Koirala}}, \bibinfo {author}
  {\bibfnamefont {J.}~\bibnamefont {Moon}}, \bibinfo {author} {\bibfnamefont
  {S.}~\bibnamefont {Oh}}, \ and\ \bibinfo {author} {\bibfnamefont {N.~P.}\
  \bibnamefont {Armitage}},\ }\bibfield  {title} {\enquote {\bibinfo {title}
  {Quantized faraday and kerr rotation and axion electrodynamics of a 3d
  topological insulator},}\ }\href {\doibase 10.1126/science.aaf5541}
  {\bibfield  {journal} {\bibinfo  {journal} {Science}\ }\textbf {\bibinfo
  {volume} {354}},\ \bibinfo {pages} {1124--1127} (\bibinfo {year}
  {2016})}\BibitemShut {NoStop}%
\bibitem [{\citenamefont {Okada}\ \emph {et~al.}(2016)\citenamefont {Okada},
  \citenamefont {Takahashi}, \citenamefont {Mogi}, \citenamefont {Yoshimi},
  \citenamefont {Tsukazaki}, \citenamefont {Takahashi}, \citenamefont {Ogawa},
  \citenamefont {Kawasaki},\ and\ \citenamefont {Tokura}}]{okada2016}%
  \BibitemOpen
  \bibfield  {author} {\bibinfo {author} {\bibfnamefont {Ken~N.}\ \bibnamefont
  {Okada}}, \bibinfo {author} {\bibfnamefont {Youtarou}\ \bibnamefont
  {Takahashi}}, \bibinfo {author} {\bibfnamefont {Masataka}\ \bibnamefont
  {Mogi}}, \bibinfo {author} {\bibfnamefont {Ryutaro}\ \bibnamefont {Yoshimi}},
  \bibinfo {author} {\bibfnamefont {Atsushi}\ \bibnamefont {Tsukazaki}},
  \bibinfo {author} {\bibfnamefont {Kei~S.}\ \bibnamefont {Takahashi}},
  \bibinfo {author} {\bibfnamefont {Naoki}\ \bibnamefont {Ogawa}}, \bibinfo
  {author} {\bibfnamefont {Masashi}\ \bibnamefont {Kawasaki}}, \ and\ \bibinfo
  {author} {\bibfnamefont {Yoshinori}\ \bibnamefont {Tokura}},\ }\bibfield
  {title} {\enquote {\bibinfo {title} {Terahertz spectroscopy on faraday and
  kerr rotations in a quantum anomalous hall state},}\ }\href {\doibase
  10.1038/ncomms12245} {\bibfield  {journal} {\bibinfo  {journal} {Nat.
  Commun.}\ }\textbf {\bibinfo {volume} {7}},\ \bibinfo {pages} {12245}
  (\bibinfo {year} {2016})}\BibitemShut {NoStop}%
\bibitem [{\citenamefont {Dziom}\ \emph {et~al.}(2017)\citenamefont {Dziom},
  \citenamefont {Shuvaev}, \citenamefont {Pimenov}, \citenamefont {Astakhov},
  \citenamefont {Ames}, \citenamefont {Bendias}, \citenamefont {B\"{o}ttcher},
  \citenamefont {Tkachov}, \citenamefont {Hankiewicz}, \citenamefont
  {Br\"{u}ne}, \citenamefont {Buhmann},\ and\ \citenamefont
  {Molenkamp}}]{dziom2017}%
  \BibitemOpen
  \bibfield  {author} {\bibinfo {author} {\bibfnamefont {V.}~\bibnamefont
  {Dziom}}, \bibinfo {author} {\bibfnamefont {A.}~\bibnamefont {Shuvaev}},
  \bibinfo {author} {\bibfnamefont {A.}~\bibnamefont {Pimenov}}, \bibinfo
  {author} {\bibfnamefont {G.~V.}\ \bibnamefont {Astakhov}}, \bibinfo {author}
  {\bibfnamefont {C.}~\bibnamefont {Ames}}, \bibinfo {author} {\bibfnamefont
  {K.}~\bibnamefont {Bendias}}, \bibinfo {author} {\bibfnamefont
  {J.}~\bibnamefont {B\"{o}ttcher}}, \bibinfo {author} {\bibfnamefont
  {G.}~\bibnamefont {Tkachov}}, \bibinfo {author} {\bibfnamefont {E.~M.}\
  \bibnamefont {Hankiewicz}}, \bibinfo {author} {\bibfnamefont
  {C.}~\bibnamefont {Br\"{u}ne}}, \bibinfo {author} {\bibfnamefont
  {H.}~\bibnamefont {Buhmann}}, \ and\ \bibinfo {author} {\bibfnamefont
  {L.~W.}\ \bibnamefont {Molenkamp}},\ }\bibfield  {title} {\enquote {\bibinfo
  {title} {Observation of the universal magnetoelectric effect in a 3d
  topological insulator},}\ }\href {\doibase 10.1038/ncomms15197} {\bibfield
  {journal} {\bibinfo  {journal} {Nat. Commun.}\ }\textbf {\bibinfo {volume}
  {8}},\ \bibinfo {pages} {15197} (\bibinfo {year} {2017})}\BibitemShut
  {NoStop}%
\bibitem [{\citenamefont {Mogi}\ \emph {et~al.}()\citenamefont {Mogi},
  \citenamefont {Okamura}, \citenamefont {Kawamura}, \citenamefont {Yoshimi},
  \citenamefont {Yasuda}, \citenamefont {Tsukazaki}, \citenamefont {Takahashi},
  \citenamefont {Morimoto}, \citenamefont {Nagaosa}, \citenamefont {Kawasaki},
  \citenamefont {Takahashi},\ and\ \citenamefont {Tokura}}]{mogi2021}%
  \BibitemOpen
  \bibfield  {author} {\bibinfo {author} {\bibfnamefont {M.}~\bibnamefont
  {Mogi}}, \bibinfo {author} {\bibfnamefont {Y.}~\bibnamefont {Okamura}},
  \bibinfo {author} {\bibfnamefont {M.}~\bibnamefont {Kawamura}}, \bibinfo
  {author} {\bibfnamefont {R.}~\bibnamefont {Yoshimi}}, \bibinfo {author}
  {\bibfnamefont {K.}~\bibnamefont {Yasuda}}, \bibinfo {author} {\bibfnamefont
  {A.}~\bibnamefont {Tsukazaki}}, \bibinfo {author} {\bibfnamefont {K.~S.}\
  \bibnamefont {Takahashi}}, \bibinfo {author} {\bibfnamefont {T.}~\bibnamefont
  {Morimoto}}, \bibinfo {author} {\bibfnamefont {N.}~\bibnamefont {Nagaosa}},
  \bibinfo {author} {\bibfnamefont {M.}~\bibnamefont {Kawasaki}}, \bibinfo
  {author} {\bibfnamefont {Y.}~\bibnamefont {Takahashi}}, \ and\ \bibinfo
  {author} {\bibfnamefont {Y.}~\bibnamefont {Tokura}},\ }\bibfield  {title}
  {\enquote {\bibinfo {title} {Experimental signature of parity anomaly in
  semi-magnetic topological insulator},}\ }\href@noop {} {\bibinfo  {journal}
  {arXiv:2105.04127}\ }\BibitemShut {NoStop}%
\bibitem [{\citenamefont {Qi}\ \emph {et~al.}(2009)\citenamefont {Qi},
  \citenamefont {Li}, \citenamefont {Zang},\ and\ \citenamefont
  {Zhang}}]{qi2009b}%
  \BibitemOpen
\bibfield  {journal} {  }\bibfield  {author} {\bibinfo {author} {\bibfnamefont
  {Xiao-Liang}\ \bibnamefont {Qi}}, \bibinfo {author} {\bibfnamefont {Rundong}\
  \bibnamefont {Li}}, \bibinfo {author} {\bibfnamefont {Jiadong}\ \bibnamefont
  {Zang}}, \ and\ \bibinfo {author} {\bibfnamefont {Shou-Cheng}\ \bibnamefont
  {Zhang}},\ }\bibfield  {title} {\enquote {\bibinfo {title} {Seeing the
  magnetic monopole through the mirror of topological surface states},}\ }\href
  {\doibase 10.1126/ science.1167747} {\bibfield  {journal} {\bibinfo
  {journal} {Science}\ }\textbf {\bibinfo {volume} {323}},\ \bibinfo {pages}
  {1184} (\bibinfo {year} {2009})}\BibitemShut {NoStop}%
\bibitem [{\citenamefont {Fu}\ and\ \citenamefont {Kane}(2008)}]{fu2008}%
  \BibitemOpen
  \bibfield  {author} {\bibinfo {author} {\bibfnamefont {Liang}\ \bibnamefont
  {Fu}}\ and\ \bibinfo {author} {\bibfnamefont {C.~L.}\ \bibnamefont {Kane}},\
  }\bibfield  {title} {\enquote {\bibinfo {title} {Superconducting proximity
  effect and majorana fermions at the surface of a topological insulator},}\
  }\href {\doibase 10.1103/PhysRevLett.100.096407} {\bibfield  {journal}
  {\bibinfo  {journal} {Phys. Rev. Lett.}\ }\textbf {\bibinfo {volume} {100}},\
  \bibinfo {pages} {096407} (\bibinfo {year} {2008})}\BibitemShut {NoStop}%
\bibitem [{\citenamefont {Hsieh}\ \emph {et~al.}(2011)\citenamefont {Hsieh},
  \citenamefont {Mahmood}, \citenamefont {McIver}, \citenamefont {Gardner},
  \citenamefont {Lee},\ and\ \citenamefont {Gedik}}]{hsieh2011}%
  \BibitemOpen
  \bibfield  {author} {\bibinfo {author} {\bibfnamefont {D.}~\bibnamefont
  {Hsieh}}, \bibinfo {author} {\bibfnamefont {F.}~\bibnamefont {Mahmood}},
  \bibinfo {author} {\bibfnamefont {J.~W.}\ \bibnamefont {McIver}}, \bibinfo
  {author} {\bibfnamefont {D.~R.}\ \bibnamefont {Gardner}}, \bibinfo {author}
  {\bibfnamefont {Y.~S.}\ \bibnamefont {Lee}}, \ and\ \bibinfo {author}
  {\bibfnamefont {N.}~\bibnamefont {Gedik}},\ }\bibfield  {title} {\enquote
  {\bibinfo {title} {Selective probing of photoinduced charge and spin dynamics
  in the bulk and surface of a topological insulator},}\ }\href {\doibase
  10.1103/PhysRevLett.107.077401} {\bibfield  {journal} {\bibinfo  {journal}
  {Phys. Rev. Lett.}\ }\textbf {\bibinfo {volume} {107}},\ \bibinfo {pages}
  {077401} (\bibinfo {year} {2011})}\BibitemShut {NoStop}%
\bibitem [{\citenamefont {Wang}\ \emph {et~al.}(2012)\citenamefont {Wang},
  \citenamefont {Hsieh}, \citenamefont {Sie}, \citenamefont {Steinberg},
  \citenamefont {Gardner}, \citenamefont {Lee}, \citenamefont
  {Jarillo-Herrero},\ and\ \citenamefont {Gedik}}]{wangyh2012}%
  \BibitemOpen
  \bibfield  {author} {\bibinfo {author} {\bibfnamefont {Y.~H.}\ \bibnamefont
  {Wang}}, \bibinfo {author} {\bibfnamefont {D.}~\bibnamefont {Hsieh}},
  \bibinfo {author} {\bibfnamefont {E.~J.}\ \bibnamefont {Sie}}, \bibinfo
  {author} {\bibfnamefont {H.}~\bibnamefont {Steinberg}}, \bibinfo {author}
  {\bibfnamefont {D.~R.}\ \bibnamefont {Gardner}}, \bibinfo {author}
  {\bibfnamefont {Y.~S.}\ \bibnamefont {Lee}}, \bibinfo {author} {\bibfnamefont
  {P.}~\bibnamefont {Jarillo-Herrero}}, \ and\ \bibinfo {author} {\bibfnamefont
  {N.}~\bibnamefont {Gedik}},\ }\bibfield  {title} {\enquote {\bibinfo {title}
  {Measurement of intrinsic dirac fermion cooling on the surface of the
  topological insulator ${\mathrm{bi}}_{2}{\mathrm{se}}_{3}$ using
  time-resolved and angle-resolved photoemission spectroscopy},}\ }\href
  {\doibase 10.1103/PhysRevLett.109.127401} {\bibfield  {journal} {\bibinfo
  {journal} {Phys. Rev. Lett.}\ }\textbf {\bibinfo {volume} {109}},\ \bibinfo
  {pages} {127401} (\bibinfo {year} {2012})}\BibitemShut {NoStop}%
\bibitem [{\citenamefont {Niesner}\ \emph {et~al.}(2012)\citenamefont
  {Niesner}, \citenamefont {Fauster}, \citenamefont {Eremeev}, \citenamefont
  {Menshchikova}, \citenamefont {Koroteev}, \citenamefont {Protogenov},
  \citenamefont {Chulkov}, \citenamefont {Tereshchenko}, \citenamefont {Kokh},
  \citenamefont {Alekperov}, \citenamefont {Nadjafov},\ and\ \citenamefont
  {Mamedov}}]{niesner2012}%
  \BibitemOpen
  \bibfield  {author} {\bibinfo {author} {\bibfnamefont {D.}~\bibnamefont
  {Niesner}}, \bibinfo {author} {\bibfnamefont {Th.}\ \bibnamefont {Fauster}},
  \bibinfo {author} {\bibfnamefont {S.~V.}\ \bibnamefont {Eremeev}}, \bibinfo
  {author} {\bibfnamefont {T.~V.}\ \bibnamefont {Menshchikova}}, \bibinfo
  {author} {\bibfnamefont {Yu.~M.}\ \bibnamefont {Koroteev}}, \bibinfo {author}
  {\bibfnamefont {A.~P.}\ \bibnamefont {Protogenov}}, \bibinfo {author}
  {\bibfnamefont {E.~V.}\ \bibnamefont {Chulkov}}, \bibinfo {author}
  {\bibfnamefont {O.~E.}\ \bibnamefont {Tereshchenko}}, \bibinfo {author}
  {\bibfnamefont {K.~A.}\ \bibnamefont {Kokh}}, \bibinfo {author}
  {\bibfnamefont {O.}~\bibnamefont {Alekperov}}, \bibinfo {author}
  {\bibfnamefont {A.}~\bibnamefont {Nadjafov}}, \ and\ \bibinfo {author}
  {\bibfnamefont {N.}~\bibnamefont {Mamedov}},\ }\bibfield  {title} {\enquote
  {\bibinfo {title} {Unoccupied topological states on bismuth chalcogenides},}\
  }\href {\doibase 10.1103/PhysRevB.86.205403} {\bibfield  {journal} {\bibinfo
  {journal} {Phys. Rev. B}\ }\textbf {\bibinfo {volume} {86}},\ \bibinfo
  {pages} {205403} (\bibinfo {year} {2012})}\BibitemShut {NoStop}%
\bibitem [{\citenamefont {Sobota}\ \emph {et~al.}(2013)\citenamefont {Sobota},
  \citenamefont {Yang}, \citenamefont {Kemper}, \citenamefont {Lee},
  \citenamefont {Schmitt}, \citenamefont {Li}, \citenamefont {Moore},
  \citenamefont {Analytis}, \citenamefont {Fisher}, \citenamefont {Kirchmann},
  \citenamefont {Devereaux},\ and\ \citenamefont {Shen}}]{sobota2013}%
  \BibitemOpen
  \bibfield  {author} {\bibinfo {author} {\bibfnamefont {J.~A.}\ \bibnamefont
  {Sobota}}, \bibinfo {author} {\bibfnamefont {S.-L.}\ \bibnamefont {Yang}},
  \bibinfo {author} {\bibfnamefont {A.~F.}\ \bibnamefont {Kemper}}, \bibinfo
  {author} {\bibfnamefont {J.~J.}\ \bibnamefont {Lee}}, \bibinfo {author}
  {\bibfnamefont {F.~T.}\ \bibnamefont {Schmitt}}, \bibinfo {author}
  {\bibfnamefont {W.}~\bibnamefont {Li}}, \bibinfo {author} {\bibfnamefont
  {R.~G.}\ \bibnamefont {Moore}}, \bibinfo {author} {\bibfnamefont {J.~G.}\
  \bibnamefont {Analytis}}, \bibinfo {author} {\bibfnamefont {I.~R.}\
  \bibnamefont {Fisher}}, \bibinfo {author} {\bibfnamefont {P.~S.}\
  \bibnamefont {Kirchmann}}, \bibinfo {author} {\bibfnamefont {T.~P.}\
  \bibnamefont {Devereaux}}, \ and\ \bibinfo {author} {\bibfnamefont {Z.-X.}\
  \bibnamefont {Shen}},\ }\bibfield  {title} {\enquote {\bibinfo {title}
  {Direct optical coupling to an unoccupied dirac surface state in the
  topological insulator ${\mathrm{bi}}_{2}{\mathrm{se}}_{3}$},}\ }\href
  {\doibase 10.1103/PhysRevLett.111.136802} {\bibfield  {journal} {\bibinfo
  {journal} {Phys. Rev. Lett.}\ }\textbf {\bibinfo {volume} {111}},\ \bibinfo
  {pages} {136802} (\bibinfo {year} {2013})}\BibitemShut {NoStop}%
\bibitem [{\citenamefont {Patankar}\ \emph {et~al.}(2015)\citenamefont
  {Patankar}, \citenamefont {Hinton}, \citenamefont {Griesmar}, \citenamefont
  {Orenstein}, \citenamefont {Dodge}, \citenamefont {Kou}, \citenamefont {Pan},
  \citenamefont {Wang}, \citenamefont {Bestwick}, \citenamefont {Fox},
  \citenamefont {Goldhaber-Gordon}, \citenamefont {Wang},\ and\ \citenamefont
  {Zhang}}]{patankar2015}%
  \BibitemOpen
  \bibfield  {author} {\bibinfo {author} {\bibfnamefont {Shreyas}\ \bibnamefont
  {Patankar}}, \bibinfo {author} {\bibfnamefont {J.~P.}\ \bibnamefont
  {Hinton}}, \bibinfo {author} {\bibfnamefont {Joel}\ \bibnamefont {Griesmar}},
  \bibinfo {author} {\bibfnamefont {J.}~\bibnamefont {Orenstein}}, \bibinfo
  {author} {\bibfnamefont {J.~S.}\ \bibnamefont {Dodge}}, \bibinfo {author}
  {\bibfnamefont {Xufeng}\ \bibnamefont {Kou}}, \bibinfo {author}
  {\bibfnamefont {Lei}\ \bibnamefont {Pan}}, \bibinfo {author} {\bibfnamefont
  {Kang~L.}\ \bibnamefont {Wang}}, \bibinfo {author} {\bibfnamefont {A.~J.}\
  \bibnamefont {Bestwick}}, \bibinfo {author} {\bibfnamefont {E.~J.}\
  \bibnamefont {Fox}}, \bibinfo {author} {\bibfnamefont {D.}~\bibnamefont
  {Goldhaber-Gordon}}, \bibinfo {author} {\bibfnamefont {Jing}\ \bibnamefont
  {Wang}}, \ and\ \bibinfo {author} {\bibfnamefont {Shou-Cheng}\ \bibnamefont
  {Zhang}},\ }\bibfield  {title} {\enquote {\bibinfo {title} {Resonant
  magneto-optic kerr effect in the magnetic topological insulator
  $\mathrm{Cr}:({\mathrm{sb}}_{x},{\mathrm{bi}}_{1\ensuremath{-}x}{)}_{2}{\mathrm{te}}_{3}$},}\
  }\href {\doibase 10.1103/PhysRevB.92.214440} {\bibfield  {journal} {\bibinfo
  {journal} {Phys. Rev. B}\ }\textbf {\bibinfo {volume} {92}},\ \bibinfo
  {pages} {214440} (\bibinfo {year} {2015})}\BibitemShut {NoStop}%
\bibitem [{\citenamefont {Zhang}\ \emph {et~al.}(2019)\citenamefont {Zhang},
  \citenamefont {Shi}, \citenamefont {Zhu}, \citenamefont {Xing}, \citenamefont
  {Zhang},\ and\ \citenamefont {Wang}}]{zhang2019}%
  \BibitemOpen
  \bibfield  {author} {\bibinfo {author} {\bibfnamefont {Dongqin}\ \bibnamefont
  {Zhang}}, \bibinfo {author} {\bibfnamefont {Minji}\ \bibnamefont {Shi}},
  \bibinfo {author} {\bibfnamefont {Tongshuai}\ \bibnamefont {Zhu}}, \bibinfo
  {author} {\bibfnamefont {Dingyu}\ \bibnamefont {Xing}}, \bibinfo {author}
  {\bibfnamefont {Haijun}\ \bibnamefont {Zhang}}, \ and\ \bibinfo {author}
  {\bibfnamefont {Jing}\ \bibnamefont {Wang}},\ }\bibfield  {title} {\enquote
  {\bibinfo {title} {Topological axion states in the magnetic insulator
  ${\mathrm{mnbi}}_{2}{\mathrm{te}}_{4}$ with the quantized magnetoelectric
  effect},}\ }\href {\doibase 10.1103/PhysRevLett.122.206401} {\bibfield
  {journal} {\bibinfo  {journal} {Phys. Rev. Lett.}\ }\textbf {\bibinfo
  {volume} {122}},\ \bibinfo {pages} {206401} (\bibinfo {year}
  {2019})}\BibitemShut {NoStop}%
\bibitem [{\citenamefont {Li}\ \emph {et~al.}(2019{\natexlab{a}})\citenamefont
  {Li}, \citenamefont {Li}, \citenamefont {Du}, \citenamefont {Wang},
  \citenamefont {Gu}, \citenamefont {Zhang}, \citenamefont {He}, \citenamefont
  {Duan},\ and\ \citenamefont {Xu}}]{li2019}%
  \BibitemOpen
  \bibfield  {author} {\bibinfo {author} {\bibfnamefont {Jiaheng}\ \bibnamefont
  {Li}}, \bibinfo {author} {\bibfnamefont {Yang}\ \bibnamefont {Li}}, \bibinfo
  {author} {\bibfnamefont {Shiqiao}\ \bibnamefont {Du}}, \bibinfo {author}
  {\bibfnamefont {Zun}\ \bibnamefont {Wang}}, \bibinfo {author} {\bibfnamefont
  {Bing-Lin}\ \bibnamefont {Gu}}, \bibinfo {author} {\bibfnamefont
  {Shou-Cheng}\ \bibnamefont {Zhang}}, \bibinfo {author} {\bibfnamefont
  {Ke}~\bibnamefont {He}}, \bibinfo {author} {\bibfnamefont {Wenhui}\
  \bibnamefont {Duan}}, \ and\ \bibinfo {author} {\bibfnamefont {Yong}\
  \bibnamefont {Xu}},\ }\bibfield  {title} {\enquote {\bibinfo {title}
  {Intrinsic magnetic topological insulators in van der waals layered
  mnbi2te4-family materials},}\ }\href {\doibase 10.1126/sciadv.aaw5685}
  {\bibfield  {journal} {\bibinfo  {journal} {Sci. Adv.}\ }\textbf {\bibinfo
  {volume} {5}},\ \bibinfo {pages} {eaaw5685} (\bibinfo {year}
  {2019}{\natexlab{a}})}\BibitemShut {NoStop}%
\bibitem [{\citenamefont {Gong}\ \emph {et~al.}(2019)\citenamefont {Gong},
  \citenamefont {Guo}, \citenamefont {Li}, \citenamefont {Zhu}, \citenamefont
  {Liao}, \citenamefont {Liu}, \citenamefont {Zhang}, \citenamefont {Gu},
  \citenamefont {Tang}, \citenamefont {Feng}, \citenamefont {Zhang},
  \citenamefont {Li}, \citenamefont {Song}, \citenamefont {Wang}, \citenamefont
  {Yu}, \citenamefont {Chen}, \citenamefont {Wang}, \citenamefont {Yao},
  \citenamefont {Duan}, \citenamefont {Xu}, \citenamefont {Zhang},
  \citenamefont {Ma}, \citenamefont {Xue},\ and\ \citenamefont
  {He}}]{gong2019}%
  \BibitemOpen
  \bibfield  {author} {\bibinfo {author} {\bibfnamefont {Yan}\ \bibnamefont
  {Gong}}, \bibinfo {author} {\bibfnamefont {Jingwen}\ \bibnamefont {Guo}},
  \bibinfo {author} {\bibfnamefont {Jiaheng}\ \bibnamefont {Li}}, \bibinfo
  {author} {\bibfnamefont {Kejing}\ \bibnamefont {Zhu}}, \bibinfo {author}
  {\bibfnamefont {Menghan}\ \bibnamefont {Liao}}, \bibinfo {author}
  {\bibfnamefont {Xiaozhi}\ \bibnamefont {Liu}}, \bibinfo {author}
  {\bibfnamefont {Qinghua}\ \bibnamefont {Zhang}}, \bibinfo {author}
  {\bibfnamefont {Lin}\ \bibnamefont {Gu}}, \bibinfo {author} {\bibfnamefont
  {Lin}\ \bibnamefont {Tang}}, \bibinfo {author} {\bibfnamefont {Xiao}\
  \bibnamefont {Feng}}, \bibinfo {author} {\bibfnamefont {Ding}\ \bibnamefont
  {Zhang}}, \bibinfo {author} {\bibfnamefont {Wei}\ \bibnamefont {Li}},
  \bibinfo {author} {\bibfnamefont {Canli}\ \bibnamefont {Song}}, \bibinfo
  {author} {\bibfnamefont {Lili}\ \bibnamefont {Wang}}, \bibinfo {author}
  {\bibfnamefont {Pu}~\bibnamefont {Yu}}, \bibinfo {author} {\bibfnamefont
  {Xi}~\bibnamefont {Chen}}, \bibinfo {author} {\bibfnamefont {Yayu}\
  \bibnamefont {Wang}}, \bibinfo {author} {\bibfnamefont {Hong}\ \bibnamefont
  {Yao}}, \bibinfo {author} {\bibfnamefont {Wenhui}\ \bibnamefont {Duan}},
  \bibinfo {author} {\bibfnamefont {Yong}\ \bibnamefont {Xu}}, \bibinfo
  {author} {\bibfnamefont {Shou-Cheng}\ \bibnamefont {Zhang}}, \bibinfo
  {author} {\bibfnamefont {Xucun}\ \bibnamefont {Ma}}, \bibinfo {author}
  {\bibfnamefont {Qi-Kun}\ \bibnamefont {Xue}}, \ and\ \bibinfo {author}
  {\bibfnamefont {Ke}~\bibnamefont {He}},\ }\bibfield  {title} {\enquote
  {\bibinfo {title} {Experimental realization of an intrinsic magnetic
  topological insulator},}\ }\href {\doibase 10.1088/0256-307X/36/7/076801}
  {\bibfield  {journal} {\bibinfo  {journal} {Chin. Phys. Lett.}\ }\textbf
  {\bibinfo {volume} {36}},\ \bibinfo {eid} {076801} (\bibinfo {year}
  {2019})}\BibitemShut {NoStop}%
\bibitem [{\citenamefont {{Otrokov}}\ \emph {et~al.}(2019)\citenamefont
  {{Otrokov}}, \citenamefont {{Klimovskikh}}, \citenamefont {{Bentmann}},
  \citenamefont {{Zeugner}}, \citenamefont {{Aliev}}, \citenamefont {{Gass}},
  \citenamefont {{Wolter}}, \citenamefont {{Koroleva}}, \citenamefont
  {{Estyunin}}, \citenamefont {{Shikin}}, \citenamefont {{Blanco-Rey}},
  \citenamefont {{Hoffmann}}, \citenamefont {{Vyazovskaya}}, \citenamefont
  {{Eremeev}}, \citenamefont {{Koroteev}}, \citenamefont {{Amiraslanov}},
  \citenamefont {{Babanly}}, \citenamefont {{Mamedov}}, \citenamefont
  {{Abdullayev}}, \citenamefont {{Zverev}}, \citenamefont {{B{\"u}chner}},
  \citenamefont {{Schwier}}, \citenamefont {{Kumar}}, \citenamefont {{Kimura}},
  \citenamefont {{Petaccia}}, \citenamefont {{Di Santo}}, \citenamefont
  {{Vidal}}, \citenamefont {{Schatz}}, \citenamefont {{Ki{\ss}ner}},
  \citenamefont {{Min}}, \citenamefont {{Moser}}, \citenamefont {{Peixoto}},
  \citenamefont {{Reinert}}, \citenamefont {{Ernst}}, \citenamefont
  {{Echenique}}, \citenamefont {{Isaeva}},\ and\ \citenamefont
  {{Chulkov}}}]{otrokov2019}%
  \BibitemOpen
  \bibfield  {author} {\bibinfo {author} {\bibfnamefont {Mikhail~M.}\
  \bibnamefont {{Otrokov}}}, \bibinfo {author} {\bibfnamefont {Ilya~I.}\
  \bibnamefont {{Klimovskikh}}}, \bibinfo {author} {\bibfnamefont {Hendrik}\
  \bibnamefont {{Bentmann}}}, \bibinfo {author} {\bibfnamefont {Alexander}\
  \bibnamefont {{Zeugner}}}, \bibinfo {author} {\bibfnamefont {Ziya~S.}\
  \bibnamefont {{Aliev}}}, \bibinfo {author} {\bibfnamefont {Sebastian}\
  \bibnamefont {{Gass}}}, \bibinfo {author} {\bibfnamefont {Anja U.~B.}\
  \bibnamefont {{Wolter}}}, \bibinfo {author} {\bibfnamefont {Alexand ra~V.}\
  \bibnamefont {{Koroleva}}}, \bibinfo {author} {\bibfnamefont {Dmitry}\
  \bibnamefont {{Estyunin}}}, \bibinfo {author} {\bibfnamefont {Alexander~M.}\
  \bibnamefont {{Shikin}}}, \bibinfo {author} {\bibfnamefont {Mar{\'\i}a}\
  \bibnamefont {{Blanco-Rey}}}, \bibinfo {author} {\bibfnamefont {Martin}\
  \bibnamefont {{Hoffmann}}}, \bibinfo {author} {\bibfnamefont {Alexand
  ra~Yu.}\ \bibnamefont {{Vyazovskaya}}}, \bibinfo {author} {\bibfnamefont
  {Sergey~V.}\ \bibnamefont {{Eremeev}}}, \bibinfo {author} {\bibfnamefont
  {Yury~M.}\ \bibnamefont {{Koroteev}}}, \bibinfo {author} {\bibfnamefont
  {Imamaddin~R.}\ \bibnamefont {{Amiraslanov}}}, \bibinfo {author}
  {\bibfnamefont {Mahammad~B.}\ \bibnamefont {{Babanly}}}, \bibinfo {author}
  {\bibfnamefont {Nazim~T.}\ \bibnamefont {{Mamedov}}}, \bibinfo {author}
  {\bibfnamefont {Nadir~A.}\ \bibnamefont {{Abdullayev}}}, \bibinfo {author}
  {\bibfnamefont {Vladimir~N.}\ \bibnamefont {{Zverev}}}, \bibinfo {author}
  {\bibfnamefont {Bernd}\ \bibnamefont {{B{\"u}chner}}}, \bibinfo {author}
  {\bibfnamefont {Eike~F.}\ \bibnamefont {{Schwier}}}, \bibinfo {author}
  {\bibfnamefont {Shiv}\ \bibnamefont {{Kumar}}}, \bibinfo {author}
  {\bibfnamefont {Akio}\ \bibnamefont {{Kimura}}}, \bibinfo {author}
  {\bibfnamefont {Luca}\ \bibnamefont {{Petaccia}}}, \bibinfo {author}
  {\bibfnamefont {Giovanni}\ \bibnamefont {{Di Santo}}}, \bibinfo {author}
  {\bibfnamefont {Raphael~C.}\ \bibnamefont {{Vidal}}}, \bibinfo {author}
  {\bibfnamefont {Sonja}\ \bibnamefont {{Schatz}}}, \bibinfo {author}
  {\bibfnamefont {Katharina}\ \bibnamefont {{Ki{\ss}ner}}}, \bibinfo {author}
  {\bibfnamefont {Chul-Hee}\ \bibnamefont {{Min}}}, \bibinfo {author}
  {\bibfnamefont {Simon~K.}\ \bibnamefont {{Moser}}}, \bibinfo {author}
  {\bibfnamefont {Thiago R.~F.}\ \bibnamefont {{Peixoto}}}, \bibinfo {author}
  {\bibfnamefont {Friedrich}\ \bibnamefont {{Reinert}}}, \bibinfo {author}
  {\bibfnamefont {Arthur}\ \bibnamefont {{Ernst}}}, \bibinfo {author}
  {\bibfnamefont {Pedro~M.}\ \bibnamefont {{Echenique}}}, \bibinfo {author}
  {\bibfnamefont {Anna}\ \bibnamefont {{Isaeva}}}, \ and\ \bibinfo {author}
  {\bibfnamefont {Evgueni~V.}\ \bibnamefont {{Chulkov}}},\ }\bibfield  {title}
  {\enquote {\bibinfo {title} {Prediction and observation of an
  antiferromagnetic topological insulator},}\ }\href {\doibase
  10.1038/s41586-019-1840-9} {\bibfield  {journal} {\bibinfo  {journal}
  {Nature}\ }\textbf {\bibinfo {volume} {576}},\ \bibinfo {pages} {416--422}
  (\bibinfo {year} {2019})}\BibitemShut {NoStop}%
\bibitem [{\citenamefont {Hao}\ \emph {et~al.}(2019)\citenamefont {Hao},
  \citenamefont {Liu}, \citenamefont {Feng}, \citenamefont {Ma}, \citenamefont
  {Schwier}, \citenamefont {Arita}, \citenamefont {Kumar}, \citenamefont {Hu},
  \citenamefont {Lu}, \citenamefont {Zeng}, \citenamefont {Wang}, \citenamefont
  {Hao}, \citenamefont {Sun}, \citenamefont {Zhang}, \citenamefont {Mei},
  \citenamefont {Ni}, \citenamefont {Wu}, \citenamefont {Shimada},
  \citenamefont {Chen}, \citenamefont {Liu},\ and\ \citenamefont
  {Liu}}]{hao2019}%
  \BibitemOpen
  \bibfield  {author} {\bibinfo {author} {\bibfnamefont {Yu-Jie}\ \bibnamefont
  {Hao}}, \bibinfo {author} {\bibfnamefont {Pengfei}\ \bibnamefont {Liu}},
  \bibinfo {author} {\bibfnamefont {Yue}\ \bibnamefont {Feng}}, \bibinfo
  {author} {\bibfnamefont {Xiao-Ming}\ \bibnamefont {Ma}}, \bibinfo {author}
  {\bibfnamefont {Eike~F.}\ \bibnamefont {Schwier}}, \bibinfo {author}
  {\bibfnamefont {Masashi}\ \bibnamefont {Arita}}, \bibinfo {author}
  {\bibfnamefont {Shiv}\ \bibnamefont {Kumar}}, \bibinfo {author}
  {\bibfnamefont {Chaowei}\ \bibnamefont {Hu}}, \bibinfo {author}
  {\bibfnamefont {Rui'e}\ \bibnamefont {Lu}}, \bibinfo {author} {\bibfnamefont
  {Meng}\ \bibnamefont {Zeng}}, \bibinfo {author} {\bibfnamefont {Yuan}\
  \bibnamefont {Wang}}, \bibinfo {author} {\bibfnamefont {Zhanyang}\
  \bibnamefont {Hao}}, \bibinfo {author} {\bibfnamefont {Hong-Yi}\ \bibnamefont
  {Sun}}, \bibinfo {author} {\bibfnamefont {Ke}~\bibnamefont {Zhang}}, \bibinfo
  {author} {\bibfnamefont {Jiawei}\ \bibnamefont {Mei}}, \bibinfo {author}
  {\bibfnamefont {Ni}~\bibnamefont {Ni}}, \bibinfo {author} {\bibfnamefont
  {Liusuo}\ \bibnamefont {Wu}}, \bibinfo {author} {\bibfnamefont {Kenya}\
  \bibnamefont {Shimada}}, \bibinfo {author} {\bibfnamefont {Chaoyu}\
  \bibnamefont {Chen}}, \bibinfo {author} {\bibfnamefont {Qihang}\ \bibnamefont
  {Liu}}, \ and\ \bibinfo {author} {\bibfnamefont {Chang}\ \bibnamefont
  {Liu}},\ }\bibfield  {title} {\enquote {\bibinfo {title} {Gapless surface
  dirac cone in antiferromagnetic topological insulator
  ${\mathrm{mnbi}}_{2}{\mathrm{te}}_{4}$},}\ }\href {\doibase
  10.1103/PhysRevX.9.041038} {\bibfield  {journal} {\bibinfo  {journal} {Phys.
  Rev. X}\ }\textbf {\bibinfo {volume} {9}},\ \bibinfo {pages} {041038}
  (\bibinfo {year} {2019})}\BibitemShut {NoStop}%
\bibitem [{\citenamefont {Li}\ \emph {et~al.}(2019{\natexlab{b}})\citenamefont
  {Li}, \citenamefont {Gao}, \citenamefont {Duan}, \citenamefont {Xu},
  \citenamefont {Zhu}, \citenamefont {Tian}, \citenamefont {Gao}, \citenamefont
  {Fan}, \citenamefont {Rao}, \citenamefont {Huang}, \citenamefont {Li},
  \citenamefont {Yan}, \citenamefont {Liu}, \citenamefont {Liu}, \citenamefont
  {Huang}, \citenamefont {Li}, \citenamefont {Liu}, \citenamefont {Zhang},
  \citenamefont {Zhang}, \citenamefont {Kondo}, \citenamefont {Shin},
  \citenamefont {Lei}, \citenamefont {Shi}, \citenamefont {Zhang},
  \citenamefont {Weng}, \citenamefont {Qian},\ and\ \citenamefont
  {Ding}}]{lih2019}%
  \BibitemOpen
  \bibfield  {author} {\bibinfo {author} {\bibfnamefont {Hang}\ \bibnamefont
  {Li}}, \bibinfo {author} {\bibfnamefont {Shun-Ye}\ \bibnamefont {Gao}},
  \bibinfo {author} {\bibfnamefont {Shao-Feng}\ \bibnamefont {Duan}}, \bibinfo
  {author} {\bibfnamefont {Yuan-Feng}\ \bibnamefont {Xu}}, \bibinfo {author}
  {\bibfnamefont {Ke-Jia}\ \bibnamefont {Zhu}}, \bibinfo {author}
  {\bibfnamefont {Shang-Jie}\ \bibnamefont {Tian}}, \bibinfo {author}
  {\bibfnamefont {Jia-Cheng}\ \bibnamefont {Gao}}, \bibinfo {author}
  {\bibfnamefont {Wen-Hui}\ \bibnamefont {Fan}}, \bibinfo {author}
  {\bibfnamefont {Zhi-Cheng}\ \bibnamefont {Rao}}, \bibinfo {author}
  {\bibfnamefont {Jie-Rui}\ \bibnamefont {Huang}}, \bibinfo {author}
  {\bibfnamefont {Jia-Jun}\ \bibnamefont {Li}}, \bibinfo {author}
  {\bibfnamefont {Da-Yu}\ \bibnamefont {Yan}}, \bibinfo {author} {\bibfnamefont
  {Zheng-Tai}\ \bibnamefont {Liu}}, \bibinfo {author} {\bibfnamefont
  {Wan-Ling}\ \bibnamefont {Liu}}, \bibinfo {author} {\bibfnamefont {Yao-Bo}\
  \bibnamefont {Huang}}, \bibinfo {author} {\bibfnamefont {Yu-Liang}\
  \bibnamefont {Li}}, \bibinfo {author} {\bibfnamefont {Yi}~\bibnamefont
  {Liu}}, \bibinfo {author} {\bibfnamefont {Guo-Bin}\ \bibnamefont {Zhang}},
  \bibinfo {author} {\bibfnamefont {Peng}\ \bibnamefont {Zhang}}, \bibinfo
  {author} {\bibfnamefont {Takeshi}\ \bibnamefont {Kondo}}, \bibinfo {author}
  {\bibfnamefont {Shik}\ \bibnamefont {Shin}}, \bibinfo {author} {\bibfnamefont
  {He-Chang}\ \bibnamefont {Lei}}, \bibinfo {author} {\bibfnamefont {You-Guo}\
  \bibnamefont {Shi}}, \bibinfo {author} {\bibfnamefont {Wen-Tao}\ \bibnamefont
  {Zhang}}, \bibinfo {author} {\bibfnamefont {Hong-Ming}\ \bibnamefont {Weng}},
  \bibinfo {author} {\bibfnamefont {Tian}\ \bibnamefont {Qian}}, \ and\
  \bibinfo {author} {\bibfnamefont {Hong}\ \bibnamefont {Ding}},\ }\bibfield
  {title} {\enquote {\bibinfo {title} {Dirac surface states in intrinsic
  magnetic topological insulators ${\mathrm{eusn}}_{2}{\mathrm{as}}_{2}$ and
  ${\mathrm{mnbi}}_{2n}{\mathrm{te}}_{3n+1}$},}\ }\href {\doibase
  10.1103/PhysRevX.9.041039} {\bibfield  {journal} {\bibinfo  {journal} {Phys.
  Rev. X}\ }\textbf {\bibinfo {volume} {9}},\ \bibinfo {pages} {041039}
  (\bibinfo {year} {2019}{\natexlab{b}})}\BibitemShut {NoStop}%
\bibitem [{\citenamefont {Chen}\ \emph {et~al.}(2019)\citenamefont {Chen},
  \citenamefont {Xu}, \citenamefont {Li}, \citenamefont {Li}, \citenamefont
  {Wang}, \citenamefont {Zhang}, \citenamefont {Li}, \citenamefont {Wu},
  \citenamefont {Liang}, \citenamefont {Chen}, \citenamefont {Jung},
  \citenamefont {Cacho}, \citenamefont {Mao}, \citenamefont {Liu},
  \citenamefont {Wang}, \citenamefont {Guo}, \citenamefont {Xu}, \citenamefont
  {Liu}, \citenamefont {Yang},\ and\ \citenamefont {Chen}}]{chen2019}%
  \BibitemOpen
  \bibfield  {author} {\bibinfo {author} {\bibfnamefont {Y.~J.}\ \bibnamefont
  {Chen}}, \bibinfo {author} {\bibfnamefont {L.~X.}\ \bibnamefont {Xu}},
  \bibinfo {author} {\bibfnamefont {J.~H.}\ \bibnamefont {Li}}, \bibinfo
  {author} {\bibfnamefont {Y.~W.}\ \bibnamefont {Li}}, \bibinfo {author}
  {\bibfnamefont {H.~Y.}\ \bibnamefont {Wang}}, \bibinfo {author}
  {\bibfnamefont {C.~F.}\ \bibnamefont {Zhang}}, \bibinfo {author}
  {\bibfnamefont {H.}~\bibnamefont {Li}}, \bibinfo {author} {\bibfnamefont
  {Y.}~\bibnamefont {Wu}}, \bibinfo {author} {\bibfnamefont {A.~J.}\
  \bibnamefont {Liang}}, \bibinfo {author} {\bibfnamefont {C.}~\bibnamefont
  {Chen}}, \bibinfo {author} {\bibfnamefont {S.~W.}\ \bibnamefont {Jung}},
  \bibinfo {author} {\bibfnamefont {C.}~\bibnamefont {Cacho}}, \bibinfo
  {author} {\bibfnamefont {Y.~H.}\ \bibnamefont {Mao}}, \bibinfo {author}
  {\bibfnamefont {S.}~\bibnamefont {Liu}}, \bibinfo {author} {\bibfnamefont
  {M.~X.}\ \bibnamefont {Wang}}, \bibinfo {author} {\bibfnamefont {Y.~F.}\
  \bibnamefont {Guo}}, \bibinfo {author} {\bibfnamefont {Y.}~\bibnamefont
  {Xu}}, \bibinfo {author} {\bibfnamefont {Z.~K.}\ \bibnamefont {Liu}},
  \bibinfo {author} {\bibfnamefont {L.~X.}\ \bibnamefont {Yang}}, \ and\
  \bibinfo {author} {\bibfnamefont {Y.~L.}\ \bibnamefont {Chen}},\ }\bibfield
  {title} {\enquote {\bibinfo {title} {Topological electronic structure and its
  temperature evolution in antiferromagnetic topological insulator
  ${\mathrm{mnbi}}_{2}{\mathrm{te}}_{4}$},}\ }\href {\doibase
  10.1103/PhysRevX.9.041040} {\bibfield  {journal} {\bibinfo  {journal} {Phys.
  Rev. X}\ }\textbf {\bibinfo {volume} {9}},\ \bibinfo {pages} {041040}
  (\bibinfo {year} {2019})}\BibitemShut {NoStop}%
\bibitem [{\citenamefont {Swatek}\ \emph {et~al.}(2020)\citenamefont {Swatek},
  \citenamefont {Wu}, \citenamefont {Wang}, \citenamefont {Lee}, \citenamefont
  {Schrunk}, \citenamefont {Yan},\ and\ \citenamefont {Kaminski}}]{swatek2020}%
  \BibitemOpen
  \bibfield  {author} {\bibinfo {author} {\bibfnamefont {Przemyslaw}\
  \bibnamefont {Swatek}}, \bibinfo {author} {\bibfnamefont {Yun}\ \bibnamefont
  {Wu}}, \bibinfo {author} {\bibfnamefont {Lin-Lin}\ \bibnamefont {Wang}},
  \bibinfo {author} {\bibfnamefont {Kyungchan}\ \bibnamefont {Lee}}, \bibinfo
  {author} {\bibfnamefont {Benjamin}\ \bibnamefont {Schrunk}}, \bibinfo
  {author} {\bibfnamefont {Jiaqiang}\ \bibnamefont {Yan}}, \ and\ \bibinfo
  {author} {\bibfnamefont {Adam}\ \bibnamefont {Kaminski}},\ }\bibfield
  {title} {\enquote {\bibinfo {title} {Gapless dirac surface states in the
  antiferromagnetic topological insulator
  ${\mathrm{mnbi}}_{2}{\mathrm{te}}_{4}$},}\ }\href {\doibase
  10.1103/PhysRevB.101.161109} {\bibfield  {journal} {\bibinfo  {journal}
  {Phys. Rev. B}\ }\textbf {\bibinfo {volume} {101}},\ \bibinfo {pages}
  {161109} (\bibinfo {year} {2020})}\BibitemShut {NoStop}%
\bibitem [{\citenamefont {Yan}\ \emph {et~al.}(2019)\citenamefont {Yan},
  \citenamefont {Zhang}, \citenamefont {Heitmann}, \citenamefont {Huang},
  \citenamefont {Chen}, \citenamefont {Cheng}, \citenamefont {Wu},
  \citenamefont {Vaknin}, \citenamefont {Sales},\ and\ \citenamefont
  {McQueeney}}]{yan2019}%
  \BibitemOpen
  \bibfield  {author} {\bibinfo {author} {\bibfnamefont {J.-Q.}\ \bibnamefont
  {Yan}}, \bibinfo {author} {\bibfnamefont {Q.}~\bibnamefont {Zhang}}, \bibinfo
  {author} {\bibfnamefont {T.}~\bibnamefont {Heitmann}}, \bibinfo {author}
  {\bibfnamefont {Z.}~\bibnamefont {Huang}}, \bibinfo {author} {\bibfnamefont
  {K.~Y.}\ \bibnamefont {Chen}}, \bibinfo {author} {\bibfnamefont {J.-G.}\
  \bibnamefont {Cheng}}, \bibinfo {author} {\bibfnamefont {W.}~\bibnamefont
  {Wu}}, \bibinfo {author} {\bibfnamefont {D.}~\bibnamefont {Vaknin}}, \bibinfo
  {author} {\bibfnamefont {B.~C.}\ \bibnamefont {Sales}}, \ and\ \bibinfo
  {author} {\bibfnamefont {R.~J.}\ \bibnamefont {McQueeney}},\ }\bibfield
  {title} {\enquote {\bibinfo {title} {Crystal growth and magnetic structure of
  ${\mathrm{mnbi}}_{2}{\mathrm{te}}_{4}$},}\ }\href {\doibase
  10.1103/PhysRevMaterials.3.064202} {\bibfield  {journal} {\bibinfo  {journal}
  {Phys. Rev. Materials}\ }\textbf {\bibinfo {volume} {3}},\ \bibinfo {pages}
  {064202} (\bibinfo {year} {2019})}\BibitemShut {NoStop}%
\bibitem [{sup()}]{supple}%
  \BibitemOpen
  \href@noop {} {}\bibinfo {note} {See Supplemental Material for technical
  details.}\BibitemShut {Stop}%
\bibitem [{\citenamefont {Mong}\ \emph {et~al.}(2010)\citenamefont {Mong},
  \citenamefont {Essin},\ and\ \citenamefont {Moore}}]{mong2010}%
  \BibitemOpen
  \bibfield  {author} {\bibinfo {author} {\bibfnamefont {Roger S.~K.}\
  \bibnamefont {Mong}}, \bibinfo {author} {\bibfnamefont {Andrew~M.}\
  \bibnamefont {Essin}}, \ and\ \bibinfo {author} {\bibfnamefont {Joel~E.}\
  \bibnamefont {Moore}},\ }\bibfield  {title} {\enquote {\bibinfo {title}
  {Antiferromagnetic topological insulators},}\ }\href {\doibase
  10.1103/PhysRevB.81.245209} {\bibfield  {journal} {\bibinfo  {journal} {Phys.
  Rev. B}\ }\textbf {\bibinfo {volume} {81}},\ \bibinfo {pages} {245209}
  (\bibinfo {year} {2010})}\BibitemShut {NoStop}%
\bibitem [{\citenamefont {Fu}\ and\ \citenamefont {Kane}(2007)}]{fu2007a}%
  \BibitemOpen
  \bibfield  {author} {\bibinfo {author} {\bibfnamefont {Liang}\ \bibnamefont
  {Fu}}\ and\ \bibinfo {author} {\bibfnamefont {C.~L.}\ \bibnamefont {Kane}},\
  }\bibfield  {title} {\enquote {\bibinfo {title} {Topological insulators with
  inversion symmetry},}\ }\href {\doibase 10.1103/PhysRevB.76.045302}
  {\bibfield  {journal} {\bibinfo  {journal} {Phys. Rev. B}\ }\textbf {\bibinfo
  {volume} {76}},\ \bibinfo {eid} {045302} (\bibinfo {year}
  {2007})}\BibitemShut {NoStop}%
\bibitem [{\citenamefont {Fu}(2009)}]{fu2009}%
  \BibitemOpen
  \bibfield  {author} {\bibinfo {author} {\bibfnamefont {Liang}\ \bibnamefont
  {Fu}},\ }\bibfield  {title} {\enquote {\bibinfo {title} {Hexagonal warping
  effects in the surface states of the topological insulator
  ${\mathrm{bi}}_{2}{\mathrm{te}}_{3}$},}\ }\href {\doibase
  10.1103/PhysRevLett.103.266801} {\bibfield  {journal} {\bibinfo  {journal}
  {Phys. Rev. Lett.}\ }\textbf {\bibinfo {volume} {103}},\ \bibinfo {pages}
  {266801} (\bibinfo {year} {2009})}\BibitemShut {NoStop}%
\bibitem [{\citenamefont {Wang}\ \emph {et~al.}(2014)\citenamefont {Wang},
  \citenamefont {Lian},\ and\ \citenamefont {Zhang}}]{wang2014a}%
  \BibitemOpen
  \bibfield  {author} {\bibinfo {author} {\bibfnamefont {Jing}\ \bibnamefont
  {Wang}}, \bibinfo {author} {\bibfnamefont {Biao}\ \bibnamefont {Lian}}, \
  and\ \bibinfo {author} {\bibfnamefont {Shou-Cheng}\ \bibnamefont {Zhang}},\
  }\bibfield  {title} {\enquote {\bibinfo {title} {Universal scaling of the
  quantum anomalous hall plateau transition},}\ }\href {\doibase
  10.1103/PhysRevB.89.085106} {\bibfield  {journal} {\bibinfo  {journal} {Phys.
  Rev. B}\ }\textbf {\bibinfo {volume} {89}},\ \bibinfo {pages} {085106}
  (\bibinfo {year} {2014})}\BibitemShut {NoStop}%
\bibitem [{\citenamefont {Wang}\ \emph {et~al.}(2015)\citenamefont {Wang},
  \citenamefont {Lian}, \citenamefont {Qi},\ and\ \citenamefont
  {Zhang}}]{wang2015b}%
  \BibitemOpen
  \bibfield  {author} {\bibinfo {author} {\bibfnamefont {Jing}\ \bibnamefont
  {Wang}}, \bibinfo {author} {\bibfnamefont {Biao}\ \bibnamefont {Lian}},
  \bibinfo {author} {\bibfnamefont {Xiao-Liang}\ \bibnamefont {Qi}}, \ and\
  \bibinfo {author} {\bibfnamefont {Shou-Cheng}\ \bibnamefont {Zhang}},\
  }\bibfield  {title} {\enquote {\bibinfo {title} {{Quantized topological
  magnetoelectric effect of the zero-plateau quantum anomalous Hall state}},}\
  }\href {\doibase 10.1103/PhysRevB.92.081107} {\bibfield  {journal} {\bibinfo
  {journal} {Phys. Rev. B}\ }\textbf {\bibinfo {volume} {92}},\ \bibinfo
  {pages} {081107} (\bibinfo {year} {2015})}\BibitemShut {NoStop}%
\bibitem [{\citenamefont {Mogi}\ \emph {et~al.}(2017)\citenamefont {Mogi},
  \citenamefont {Kawamura}, \citenamefont {Yoshimi}, \citenamefont {Tsukazaki},
  \citenamefont {Kozuka}, \citenamefont {Shirakawa}, \citenamefont {Takahashi},
  \citenamefont {Kawasaki},\ and\ \citenamefont {Tokura}}]{mogi2017}%
  \BibitemOpen
  \bibfield  {author} {\bibinfo {author} {\bibfnamefont {M.}~\bibnamefont
  {Mogi}}, \bibinfo {author} {\bibfnamefont {M.}~\bibnamefont {Kawamura}},
  \bibinfo {author} {\bibfnamefont {R.}~\bibnamefont {Yoshimi}}, \bibinfo
  {author} {\bibfnamefont {A.}~\bibnamefont {Tsukazaki}}, \bibinfo {author}
  {\bibfnamefont {Y.}~\bibnamefont {Kozuka}}, \bibinfo {author} {\bibfnamefont
  {N.}~\bibnamefont {Shirakawa}}, \bibinfo {author} {\bibfnamefont {K.~S.}\
  \bibnamefont {Takahashi}}, \bibinfo {author} {\bibfnamefont {M.}~\bibnamefont
  {Kawasaki}}, \ and\ \bibinfo {author} {\bibfnamefont {Y.}~\bibnamefont
  {Tokura}},\ }\bibfield  {title} {\enquote {\bibinfo {title} {A magnetic
  heterostructure of topological insulators as a candidate for an axion
  insulator},}\ }\href {\doibase 10.1038/nmat4855} {\bibfield  {journal}
  {\bibinfo  {journal} {Nature Mater.}\ }\textbf {\bibinfo {volume} {16}},\
  \bibinfo {pages} {516--521} (\bibinfo {year} {2017})}\BibitemShut {NoStop}%
\bibitem [{\citenamefont {Xiao}\ \emph {et~al.}(2018)\citenamefont {Xiao},
  \citenamefont {Jiang}, \citenamefont {Shin}, \citenamefont {Wang},
  \citenamefont {Wang}, \citenamefont {Zhao}, \citenamefont {Liu},
  \citenamefont {Wu}, \citenamefont {Chan}, \citenamefont {Samarth},\ and\
  \citenamefont {Chang}}]{xiao2018}%
  \BibitemOpen
  \bibfield  {author} {\bibinfo {author} {\bibfnamefont {Di}~\bibnamefont
  {Xiao}}, \bibinfo {author} {\bibfnamefont {Jue}\ \bibnamefont {Jiang}},
  \bibinfo {author} {\bibfnamefont {Jae-Ho}\ \bibnamefont {Shin}}, \bibinfo
  {author} {\bibfnamefont {Wenbo}\ \bibnamefont {Wang}}, \bibinfo {author}
  {\bibfnamefont {Fei}\ \bibnamefont {Wang}}, \bibinfo {author} {\bibfnamefont
  {Yi-Fan}\ \bibnamefont {Zhao}}, \bibinfo {author} {\bibfnamefont {Chaoxing}\
  \bibnamefont {Liu}}, \bibinfo {author} {\bibfnamefont {Weida}\ \bibnamefont
  {Wu}}, \bibinfo {author} {\bibfnamefont {Moses H.~W.}\ \bibnamefont {Chan}},
  \bibinfo {author} {\bibfnamefont {Nitin}\ \bibnamefont {Samarth}}, \ and\
  \bibinfo {author} {\bibfnamefont {Cui-Zu}\ \bibnamefont {Chang}},\ }\bibfield
   {title} {\enquote {\bibinfo {title} {{Realization of the Axion Insulator
  State in Quantum Anomalous Hall Sandwich Heterostructures}},}\ }\href
  {\doibase 10.1103/PhysRevLett.120.056801} {\bibfield  {journal} {\bibinfo
  {journal} {Phys. Rev. Lett.}\ }\textbf {\bibinfo {volume} {120}},\ \bibinfo
  {pages} {056801} (\bibinfo {year} {2018})}\BibitemShut {NoStop}%
\bibitem [{\citenamefont {Liu}\ \emph {et~al.}(2020)\citenamefont {Liu},
  \citenamefont {Wang}, \citenamefont {Li}, \citenamefont {Wu}, \citenamefont
  {Li}, \citenamefont {Li}, \citenamefont {He}, \citenamefont {Xu},
  \citenamefont {Zhang},\ and\ \citenamefont {Wang}}]{liu2020}%
  \BibitemOpen
  \bibfield  {author} {\bibinfo {author} {\bibfnamefont {Chang}\ \bibnamefont
  {Liu}}, \bibinfo {author} {\bibfnamefont {Yongchao}\ \bibnamefont {Wang}},
  \bibinfo {author} {\bibfnamefont {Hao}\ \bibnamefont {Li}}, \bibinfo {author}
  {\bibfnamefont {Yang}\ \bibnamefont {Wu}}, \bibinfo {author} {\bibfnamefont
  {Yaoxin}\ \bibnamefont {Li}}, \bibinfo {author} {\bibfnamefont {Jiaheng}\
  \bibnamefont {Li}}, \bibinfo {author} {\bibfnamefont {Ke}~\bibnamefont {He}},
  \bibinfo {author} {\bibfnamefont {Yong}\ \bibnamefont {Xu}}, \bibinfo
  {author} {\bibfnamefont {Jinsong}\ \bibnamefont {Zhang}}, \ and\ \bibinfo
  {author} {\bibfnamefont {Yayu}\ \bibnamefont {Wang}},\ }\bibfield  {title}
  {\enquote {\bibinfo {title} {Robust axion insulator and chern insulator
  phases in a two-dimensional antiferromagnetic topological insulator},}\
  }\href {\doibase 10.1038/s41563-019-0573-3} {\bibfield  {journal} {\bibinfo
  {journal} {Nature Mat.}\ }\textbf {\bibinfo {volume} {19}},\ \bibinfo {pages}
  {522--527} (\bibinfo {year} {2020})}\BibitemShut {NoStop}%
\bibitem [{\citenamefont {Nomura}\ and\ \citenamefont
  {Nagaosa}(2011)}]{nomura2011}%
  \BibitemOpen
  \bibfield  {author} {\bibinfo {author} {\bibfnamefont {Kentaro}\ \bibnamefont
  {Nomura}}\ and\ \bibinfo {author} {\bibfnamefont {Naoto}\ \bibnamefont
  {Nagaosa}},\ }\bibfield  {title} {\enquote {\bibinfo {title}
  {Surface-quantized anomalous hall current and the magnetoelectric effect in
  magnetically disordered topological insulators},}\ }\href {\doibase
  10.1103/PhysRevLett.106.166802} {\bibfield  {journal} {\bibinfo  {journal}
  {Phys. Rev. Lett.}\ }\textbf {\bibinfo {volume} {106}},\ \bibinfo {pages}
  {166802} (\bibinfo {year} {2011})}\BibitemShut {NoStop}%
\bibitem [{\citenamefont {Yu}\ \emph {et~al.}(2019)\citenamefont {Yu},
  \citenamefont {Zang},\ and\ \citenamefont {Liu}}]{yu2019}%
  \BibitemOpen
  \bibfield  {author} {\bibinfo {author} {\bibfnamefont {Jiabin}\ \bibnamefont
  {Yu}}, \bibinfo {author} {\bibfnamefont {Jiadong}\ \bibnamefont {Zang}}, \
  and\ \bibinfo {author} {\bibfnamefont {Chao-Xing}\ \bibnamefont {Liu}},\
  }\bibfield  {title} {\enquote {\bibinfo {title} {Magnetic resonance induced
  pseudoelectric field and giant current response in axion insulators},}\
  }\href {\doibase 10.1103/PhysRevB.100.075303} {\bibfield  {journal} {\bibinfo
   {journal} {Phys. Rev. B}\ }\textbf {\bibinfo {volume} {100}},\ \bibinfo
  {pages} {075303} (\bibinfo {year} {2019})}\BibitemShut {NoStop}%
\bibitem [{\citenamefont {Liu}\ and\ \citenamefont {Wang}(2020)}]{liuzc2020a}%
  \BibitemOpen
  \bibfield  {author} {\bibinfo {author} {\bibfnamefont {Zhaochen}\
  \bibnamefont {Liu}}\ and\ \bibinfo {author} {\bibfnamefont {Jing}\
  \bibnamefont {Wang}},\ }\bibfield  {title} {\enquote {\bibinfo {title}
  {Anisotropic topological magnetoelectric effect in axion insulators},}\
  }\href {\doibase 10.1103/PhysRevB.101.205130} {\bibfield  {journal} {\bibinfo
   {journal} {Phys. Rev. B}\ }\textbf {\bibinfo {volume} {101}},\ \bibinfo
  {pages} {205130} (\bibinfo {year} {2020})}\BibitemShut {NoStop}%
\bibitem [{\citenamefont {Liu}\ \emph {et~al.}()\citenamefont {Liu},
  \citenamefont {Xiao},\ and\ \citenamefont {Wang}}]{liuzc2020b}%
  \BibitemOpen
  \bibfield  {author} {\bibinfo {author} {\bibfnamefont {Zhaochen}\
  \bibnamefont {Liu}}, \bibinfo {author} {\bibfnamefont {Jiang}\ \bibnamefont
  {Xiao}}, \ and\ \bibinfo {author} {\bibfnamefont {Jing}\ \bibnamefont
  {Wang}},\ }\bibfield  {title} {\enquote {\bibinfo {title} {Dynamical
  magnetoelectric coupling in axion insulator thin films},}\ }\href@noop {}
  {\bibinfo  {journal} {arXiv:2007.09869}\ }\BibitemShut {NoStop}%
\bibitem [{\citenamefont {Liu}\ and\ \citenamefont {Zhu}(2007)}]{liu2007}%
  \BibitemOpen
\bibfield  {journal} {  }\bibfield  {author} {\bibinfo {author} {\bibfnamefont
  {Ren-Bao}\ \bibnamefont {Liu}}\ and\ \bibinfo {author} {\bibfnamefont
  {Bang-Fen}\ \bibnamefont {Zhu}},\ }\bibfield  {title} {\enquote {\bibinfo
  {title} {High-order thz-sideband generation in semiconductors},}\ }in\ \href
  {\doibase 10.1063/1.2730455} {\emph {\bibinfo {booktitle} {AIP Conference
  Proceedings}}},\ Vol.\ \bibinfo {volume} {893}\ (\bibinfo {organization}
  {American Institute of Physics},\ \bibinfo {year} {2007})\ pp.\ \bibinfo
  {pages} {1455--1456}\BibitemShut {NoStop}%
\bibitem [{\citenamefont {Zaks}\ \emph {et~al.}(2012)\citenamefont {Zaks},
  \citenamefont {Liu},\ and\ \citenamefont {Sherwin}}]{zaks2012}%
  \BibitemOpen
  \bibfield  {author} {\bibinfo {author} {\bibfnamefont {Ben}\ \bibnamefont
  {Zaks}}, \bibinfo {author} {\bibfnamefont {Ren-Bao}\ \bibnamefont {Liu}}, \
  and\ \bibinfo {author} {\bibfnamefont {Mark~S}\ \bibnamefont {Sherwin}},\
  }\bibfield  {title} {\enquote {\bibinfo {title} {Experimental observation of
  electron--hole recollisions},}\ }\href {\doibase 10.1038/nature10864}
  {\bibfield  {journal} {\bibinfo  {journal} {Nature}\ }\textbf {\bibinfo
  {volume} {483}},\ \bibinfo {pages} {580--583} (\bibinfo {year}
  {2012})}\BibitemShut {NoStop}%
\bibitem [{\citenamefont {Yang}\ and\ \citenamefont {Liu}(2013)}]{yang2013}%
  \BibitemOpen
  \bibfield  {author} {\bibinfo {author} {\bibfnamefont {Fan}\ \bibnamefont
  {Yang}}\ and\ \bibinfo {author} {\bibfnamefont {Ren-Bao}\ \bibnamefont
  {Liu}},\ }\bibfield  {title} {\enquote {\bibinfo {title} {Berry phases of
  quantum trajectories of optically excited electron--hole pairs in
  semiconductors under strong terahertz fields},}\ }\href {\doibase
  10.1088/1367-2630/15/11/115005} {\bibfield  {journal} {\bibinfo  {journal}
  {New J. Phys.}\ }\textbf {\bibinfo {volume} {15}},\ \bibinfo {pages} {115005}
  (\bibinfo {year} {2013})}\BibitemShut {NoStop}%
\bibitem [{\citenamefont {Yang}\ \emph {et~al.}(2014)\citenamefont {Yang},
  \citenamefont {Xu},\ and\ \citenamefont {Liu}}]{yang2014}%
  \BibitemOpen
  \bibfield  {author} {\bibinfo {author} {\bibfnamefont {Fan}\ \bibnamefont
  {Yang}}, \bibinfo {author} {\bibfnamefont {Xiaodong}\ \bibnamefont {Xu}}, \
  and\ \bibinfo {author} {\bibfnamefont {Ren-Bao}\ \bibnamefont {Liu}},\
  }\bibfield  {title} {\enquote {\bibinfo {title} {Giant faraday rotation
  induced by the berry phase in bilayer graphene under strong terahertz
  fields},}\ }\href {\doibase 10.1088/1367-2630/16/4/043014} {\bibfield
  {journal} {\bibinfo  {journal} {New J. Phys.}\ }\textbf {\bibinfo {volume}
  {16}},\ \bibinfo {pages} {043014} (\bibinfo {year} {2014})}\BibitemShut
  {NoStop}%
\bibitem [{\citenamefont {Banks}\ \emph {et~al.}(2017)\citenamefont {Banks},
  \citenamefont {Wu}, \citenamefont {Valovcin}, \citenamefont {Mack},
  \citenamefont {Gossard}, \citenamefont {Pfeiffer}, \citenamefont {Liu},\ and\
  \citenamefont {Sherwin}}]{banks2017}%
  \BibitemOpen
  \bibfield  {author} {\bibinfo {author} {\bibfnamefont {Hunter~B}\
  \bibnamefont {Banks}}, \bibinfo {author} {\bibfnamefont {Qile}\ \bibnamefont
  {Wu}}, \bibinfo {author} {\bibfnamefont {Darren~C}\ \bibnamefont {Valovcin}},
  \bibinfo {author} {\bibfnamefont {Shawn}\ \bibnamefont {Mack}}, \bibinfo
  {author} {\bibfnamefont {Arthur~C}\ \bibnamefont {Gossard}}, \bibinfo
  {author} {\bibfnamefont {Loren}\ \bibnamefont {Pfeiffer}}, \bibinfo {author}
  {\bibfnamefont {Ren-Bao}\ \bibnamefont {Liu}}, \ and\ \bibinfo {author}
  {\bibfnamefont {Mark~S}\ \bibnamefont {Sherwin}},\ }\bibfield  {title}
  {\enquote {\bibinfo {title} {Dynamical birefringence: electron-hole
  recollisions as probes of berry curvature},}\ }\href {\doibase
  10.1103/PhysRevX.7.041042} {\bibfield  {journal} {\bibinfo  {journal}
  {Physical Review X}\ }\textbf {\bibinfo {volume} {7}},\ \bibinfo {pages}
  {041042} (\bibinfo {year} {2017})}\BibitemShut {NoStop}%
\bibitem [{\citenamefont {Langer}\ \emph {et~al.}(2018)\citenamefont {Langer},
  \citenamefont {Schmid}, \citenamefont {Schlauderer}, \citenamefont {Gmitra},
  \citenamefont {Fabian}, \citenamefont {Nagler}, \citenamefont {Schuller},
  \citenamefont {Korn}, \citenamefont {Hawkins}, \citenamefont {Steiner},
  \citenamefont {Huttner}, \citenamefont {Koch}, \citenamefont {Kira},\ and\
  \citenamefont {Huber}}]{langer2018}%
  \BibitemOpen
  \bibfield  {author} {\bibinfo {author} {\bibfnamefont {F.}~\bibnamefont
  {Langer}}, \bibinfo {author} {\bibfnamefont {C.~P.}\ \bibnamefont {Schmid}},
  \bibinfo {author} {\bibfnamefont {S.}~\bibnamefont {Schlauderer}}, \bibinfo
  {author} {\bibfnamefont {M.}~\bibnamefont {Gmitra}}, \bibinfo {author}
  {\bibfnamefont {J.}~\bibnamefont {Fabian}}, \bibinfo {author} {\bibfnamefont
  {P.}~\bibnamefont {Nagler}}, \bibinfo {author} {\bibfnamefont
  {C.}~\bibnamefont {Schuller}}, \bibinfo {author} {\bibfnamefont
  {T.}~\bibnamefont {Korn}}, \bibinfo {author} {\bibfnamefont {P.~G.}\
  \bibnamefont {Hawkins}}, \bibinfo {author} {\bibfnamefont {J.~T.}\
  \bibnamefont {Steiner}}, \bibinfo {author} {\bibfnamefont {U.}~\bibnamefont
  {Huttner}}, \bibinfo {author} {\bibfnamefont {S.~W.}\ \bibnamefont {Koch}},
  \bibinfo {author} {\bibfnamefont {M.}~\bibnamefont {Kira}}, \ and\ \bibinfo
  {author} {\bibfnamefont {R.}~\bibnamefont {Huber}},\ }\bibfield  {title}
  {\enquote {\bibinfo {title} {Lightwave valleytronics in a monolayer of
  tungsten diselenide},}\ }\href {\doibase 10.1038/s41586-018-0013-6}
  {\bibfield  {journal} {\bibinfo  {journal} {Nature}\ }\textbf {\bibinfo
  {volume} {557}},\ \bibinfo {pages} {76--80} (\bibinfo {year}
  {2018})}\BibitemShut {NoStop}%
\bibitem [{\citenamefont {Xiao}\ \emph {et~al.}(2012)\citenamefont {Xiao},
  \citenamefont {Liu}, \citenamefont {Feng}, \citenamefont {Xu},\ and\
  \citenamefont {Yao}}]{xiao2012}%
  \BibitemOpen
  \bibfield  {author} {\bibinfo {author} {\bibfnamefont {Di}~\bibnamefont
  {Xiao}}, \bibinfo {author} {\bibfnamefont {Gui-Bin}\ \bibnamefont {Liu}},
  \bibinfo {author} {\bibfnamefont {Wanxiang}\ \bibnamefont {Feng}}, \bibinfo
  {author} {\bibfnamefont {Xiaodong}\ \bibnamefont {Xu}}, \ and\ \bibinfo
  {author} {\bibfnamefont {Wang}\ \bibnamefont {Yao}},\ }\bibfield  {title}
  {\enquote {\bibinfo {title} {Coupled spin and valley physics in monolayers of
  ${\mathrm{mos}}_{2}$ and other group-vi dichalcogenides},}\ }\href {\doibase
  10.1103/PhysRevLett.108.196802} {\bibfield  {journal} {\bibinfo  {journal}
  {Phys. Rev. Lett.}\ }\textbf {\bibinfo {volume} {108}},\ \bibinfo {pages}
  {196802} (\bibinfo {year} {2012})}\BibitemShut {NoStop}%
\bibitem [{\citenamefont {Sobota}\ \emph {et~al.}(2012)\citenamefont {Sobota},
  \citenamefont {Yang}, \citenamefont {Analytis}, \citenamefont {Chen},
  \citenamefont {Fisher}, \citenamefont {Kirchmann},\ and\ \citenamefont
  {Shen}}]{sobota2012}%
  \BibitemOpen
  \bibfield  {author} {\bibinfo {author} {\bibfnamefont {J.~A.}\ \bibnamefont
  {Sobota}}, \bibinfo {author} {\bibfnamefont {S.}~\bibnamefont {Yang}},
  \bibinfo {author} {\bibfnamefont {J.~G.}\ \bibnamefont {Analytis}}, \bibinfo
  {author} {\bibfnamefont {Y.~L.}\ \bibnamefont {Chen}}, \bibinfo {author}
  {\bibfnamefont {I.~R.}\ \bibnamefont {Fisher}}, \bibinfo {author}
  {\bibfnamefont {P.~S.}\ \bibnamefont {Kirchmann}}, \ and\ \bibinfo {author}
  {\bibfnamefont {Z.-X.}\ \bibnamefont {Shen}},\ }\bibfield  {title} {\enquote
  {\bibinfo {title} {Ultrafast optical excitation of a persistent surface-state
  population in the topological insulator
  ${\mathrm{bi}}_{2}{\mathrm{se}}_{3}$},}\ }\href {\doibase
  10.1103/PhysRevLett.108.117403} {\bibfield  {journal} {\bibinfo  {journal}
  {Phys. Rev. Lett.}\ }\textbf {\bibinfo {volume} {108}},\ \bibinfo {pages}
  {117403} (\bibinfo {year} {2012})}\BibitemShut {NoStop}%
\bibitem [{\citenamefont {Fang}\ \emph {et~al.}(2005)\citenamefont {Fang},
  \citenamefont {Terakura},\ and\ \citenamefont {Nagaosa}}]{fang2005}%
  \BibitemOpen
  \bibfield  {author} {\bibinfo {author} {\bibfnamefont {Zhong}\ \bibnamefont
  {Fang}}, \bibinfo {author} {\bibfnamefont {Kiyoyuki}\ \bibnamefont
  {Terakura}}, \ and\ \bibinfo {author} {\bibfnamefont {Naoto}\ \bibnamefont
  {Nagaosa}},\ }\bibfield  {title} {\enquote {\bibinfo {title} {Orbital physics
  in ruthenates: first-principles studies},}\ }\href {\doibase
  10.1088/1367-2630/7/1/066} {\bibfield  {journal} {\bibinfo  {journal} {New J.
  Phys.}\ }\textbf {\bibinfo {volume} {7}},\ \bibinfo {pages} {66--66}
  (\bibinfo {year} {2005})}\BibitemShut {NoStop}%
\end{thebibliography}
\end{document}